\documentclass[11pt, a4paper, twocolumn]{aastex631}
\usepackage{acronym}
\usepackage{footmisc}
\usepackage{float}
\usepackage{appendix}
\usepackage{graphicx}
\usepackage{array}
\usepackage{ragged2e}
\usepackage{amsmath}
\acrodef{GW}[GW]{gravitational wave}
\acrodef{CNN}[CNN]{convolutional neural network}
\acrodef{EM}[EM]{electromagnetic}
\acrodef{FAP}[FAP]{false alarm probability}
\acrodef{FAR}[FAR]{false alarm rate}
\acrodef{FNP}[FNP]{false negative probability}
\acrodef{TAP}[TAP]{true alarm probability}
\acrodef{LVC}[LVC]{The LIGO Scientific and Virgo Collaborations}
\acrodef{OPA}[OPA]{Open Public Alert}
\acrodef{GCN}[GCN]{Gamma-ray Coordination Network}
\acrodef{BNS}[BNS]{binary neutron star}
\acrodef{BBH}[BBH]{binary black hole merger}
\acrodef{NSBH}[NSBH]{neutron star black hole}
\acrodef{LVK}[LVK]{LIGO-Virgo-KAGRA Collaboration}
\acrodef{API}[API]{application programming interface}
\acrodef{LVC}[LVC]{LIGO Scientific and Virgo Collaboration}
\acrodef{O3}[O3]{the third observing run}
\acrodef{O2}[O2]{the second observing run}
\acrodef{O4}[O4]{fourth observing run}
\acrodef{CBC}[CBC]{compact binary coalescence}
\acrodef{MDC}[MDC]{mock data challenge}
\acrodef{ROC}[ROC]{receiver operator characteristic}
\acrodef{SNR}[SNR]{signal-to-noise ratio}
\acrodef{O1}[O1]{the first}
\acrodef{O2}[O2]{second}
\acrodef{ER15}[ER15]{the engineering run}
\acrodef{GraceDB}[GraceDB]{the gravitational wave-candidate event DataBase}
\acrodef{IOU}[IoU]{Intersection over Union}
\acrodef{YOLO}[YOLO]{You Only Look Once}
\acrodef{ROC}[ROC]{receiver operator characteristic}

\makeatletter
\newcommand\footnoteref[1]{\protected@xdef\@thefnmark{\ref{#1}}\@footnotemark}

\makeatother
\newcolumntype{Y}{>{\centering\arraybackslash}X}

\begin{document}
\title{\large \bf \texttt{\bf GSpyNetTreeS} : a machine learning solution for glitch localization in time and frequency}
\author[0009-0009-6826-4559]{Man Leong Chan}
\affil{Department of Physics and Astronomy, The University of British Columbia, Vancouver, BC V6T 1Z4, Canada}
\author[0000-0003-0316-1355]{Jess McIver}
\affil{Department of Physics and Astronomy, The University of British Columbia, Vancouver, BC V6T 1Z4, Canada}
\author{Yannick Lecoeuche}
\affil{Department of Physics and Astronomy, The University of British Columbia, Vancouver, BC V6T 1Z4, Canada}
\author{Dhatri Raghunathan}
\affil{Indian Institute of Technology Kharagpur, West Bengal, India}
\author{Sof\'ia \'Alvarez-L\'opez}
\affil{Department of Physics and Kavli Institute for Astrophysics and Space Research, Massachusetts Institute of Technology, Cambridge, 02139, USA}
\author{Julian Ding}
\affil{Department of Physics and Astronomy, The University of British Columbia, Vancouver, BC V6T 1Z4, Canada}
\author{Annudesh Liyanage}
\affil{Department of Physics and Astronomy, The University of British Columbia, Vancouver, BC V6T 1Z4, Canada}
\author{Raymond Ng}
\affil{Department of Computer Science, University of British Columbia, Vancouver, British Columbia, V6T1Z4, Canada}
\author{Heather Fong}
\affil{Department of Physics and Astronomy, The University of British Columbia, Vancouver, BC V6T 1Z4, Canada}
%\author[1]{Man Leong Chan~\orcidlink{0009-0009-6826-4559}}
%\author[1]{Jess McIver~\orcidlink{0000-0003-0316-1355}}
%\author[2,3]{Daryl Haggard~\orcidlink{0000-0001-6803-2138}}
%\author[4]{Cody Messick~\orcidlink{0000-0002-8230-3309}}
%\author[5,6]{Ashish Mahabal~\orcidlink{0000-0003-2242-0244}}
%\author[2,3]{Nayyer Raza~\orcidlink{0000-0002-8549-9124}}
%\author[7]{Patrick J. Sutton~\orcidlink{0000-0003-1614-3922}}
%\author[8,9]{Becca Ewing~\orcidlink{0000-0001-9178-5744}}
%\author[10]{Francesco Di Renzo~\orcidlink{0000-0002-5447-3810}}

%\affil[2]{Department of Physics, McGill University, 3600 rue University, Montreal, Quebec H3A2T8, Canada}
%\affil[3]{Trottier Space Institute at McGill, 3550 rue University, Montreal, Quebec H3A2A7, Canada}
%\affil[4]{University of Wisconsin-Milwaukee, Milwaukee, WI 53201, USA}
%\affil[5]{Division of Physics, Mathematics and Astronomy, California Institute of Technology, Pasadena, CA 91125, USA}
%\affil[6]{Center for Data Driven Discovery, California Institute of Technology, Pasadena, CA 91125, USA}
%\affil[7]{Gravity Exploration Institute, Cardiff University, Cardiff CF24 3AA, UK}
%\affil[8]{Department of Physics, The Pennsylvania State University, University Park, PA 16802, USA}
%\affil[9]{Institute for Gravitation and the Cosmos, The Pennsylvania State University, University Park, PA 16802, USA}
%\affil[10]{Université Lyon, Université Claude Bernard Lyon 1, CNRS, IP2I Lyon/IN2P3, UMR 5822, F-69622 Villeurbanne, France}
\begin{abstract}
Data from ground-based gravitational wave detectors are often contaminated by non-Gaussian instrumental artifacts or detector noise transients. Unbiased source property estimation relies on the ability to correctly identify and characterize these artifacts and remove them if necessary. To this end, the LIGO-Virgo-KAGRA Collaboration has implemented candidate vetting for all significant candidates to identify the presence of artifacts and assess the need for mitigation. The current candidate vetting process requires human experts to identify the frequency ranges and the time windows associated with any data quality issues present. Differences in judgment between human experts may cause inconsistency, making results difficult to reproduce across gravitational wave events. We present \texttt{GSpyNetTreeS}, an extension to \texttt{GSpyNetTree} based on the You Only Look Once algorithm, for the automatic detection, classification, and time-frequency localization of detector noise transients. As a proof of concept, we tested \texttt{GSpyNetTreeS}'s performance on the data collected by the LIGO detectors during the third observing run for gravitational waves as well as common detector glitch classes included in \texttt{GSpyNetTree}: Blip, Low frequency blip, Low frequency line and Scratchy. We also demonstrated that \texttt{GSpyNetTreeS} is capable of accurately identifying common glitch classes and capturing the frequency and time information associated with detected detector noise transients, establishing its potential as an automatic event validation tool for LIGO-Virgo-KAGRA's observing runs. 
 %An independent public version of the algorithm trained on publicly available data only will also be implemented on the Treasure Map.
\vspace{1cm}
\end{abstract}
\section{\bf Introduction}
Since the first observation of \acp{GW} from a \ac{BBH} by the LIGO detectors~\citep{LIGOScientific:2014pky} in 2015~\citep{PhysRevLett.116.061102}, detection of \acp{GW} from \acp{CBC} has become commonplace. The \ac{LVK} has reported observations of approximately $90$ \acp{GW} from the first three LIGO-Virgo-KAGRA observing runs~\citep{LIGOScientific:2018mvr, LIGOScientific:2020ibl, KAGRA:2021vkt}. With over 200 significant \ac{GW} candidates already identified in the current observing run,  \ac{O4}\footnote{https://gracedb.ligo.org/superevents/public/O4/}, the catalog of \acp{GW} is expected to quickly expand.  Thanks to improved LIGO detector sensitivity, there is also the potential to observe \acp{GW} from sources that have so far evaded detection such as core-collapse supernovae and spinning neutron stars~\citep{KAGRA:2022dwb, KAGRA:2021tnv}. 

Ground-based \ac{GW} detectors such as the LIGO, Virgo~\citep{VIRGO:2014yos} and KAGRA~\citep{Aso:2013eba} detectors are prone to a high rate of non-Gaussian and non-stationary terrestrial noises, known as glitches~\citep{LIGO:2024kkz, Davis_2021, Abbott_2020, Berger_2018}. 
%It is quite common for glitches to be present near a gravitational-wave signal. 
For example, the LIGO and Virgo detectors in \ac{O3} experienced a median glitch rate of approximately above 1 per minute, indicating a non-negligible probability ($\mathcal{O}(10\%)$) that a glitch overlaps a signal from \acp{CBC}~\citep{Davis:2022ird}. Depending on the morphology, the presence of a glitch can bias \ac{GW} source property estimates~\citep{PhysRevD.105.103021, PhysRevD.106.104017, PhysRevD.110.122002, PhysRevD.106.042006, PhysRevD.98.084016, Canton:2013joa, Powell:2018csz, Udall:2024ovp} and even mimic \ac{GW} signals, leading to false detection of \acp{GW} even after data conditioning and event ranking statistics are applied, e.g. by \ac{CBC} search pipelines GstLAL~\citep{PhysRevD.108.043004, PhysRevD.109.042008}, PyCBC~\citep{Dal_Canton_2021, Usman_2016}, MBTA~\citep{Allene_2025}, and SPIIR~\citep{PhysRevD.105.024023}. 
%The presence of a glitch can cause false positives of \acp{GW}. If a glitch occurs near a \ac{GW} signal, the glitch can bias the estimates of \ac{GW} source properties~[ref]. These methods include Gravity Spy, which is a citizen science machine learning project~[ref],

%Methods and algorithms have been developed to separate glitches from \acp{GW} in real-time to facilitate reliable real-time release of \ac{GW} candidates, including GWSkyNet~[ref], a real-time machine learning pipeline, for the identification of the origin of \ac{GW} candidates and its extension GWSkyNet-multi~[ref].  \ac{GW} search pipelines also employ signal consistency checks to reduce false positives of \acp{GW} due to glitches.  
To determine whether a \ac{GW} candidate is affected by data quality issues, 
the \ac{LVK} Collaboration has implemented candidate vetting~\citep{LIGO:2024kkz}. Candidate vetting is a process where human experts cross check a list of data quality tasks\footnote{https://detchar.docs.ligo.org/dqrtasks/index.html} for a given \ac{GW} candidate, to determine if data quality issues such as glitches are present and could affect source property estimation for the \ac{GW} candidate. For example, tasks producing results for each \ac{GW} candidate via the LIGO-Virgo Data Quality Report\footnote{https://docs.ligo.org/detchar/data-quality-report/} in O4 include \texttt{iDQ}, a supervised learning tool that detects detector noise artifacts through auxiliary channel data~\citep{PhysRevD.88.062003, Essick_2013}, \texttt{GSpyNetTree}~\citep{Alvarez-Lopez:2023dmv, Jarov:2023qpt}, a machine learning classifier for the identification of different glitch morphologies, and \texttt{GlitchFind}~\citep{Vazsonyi:2022jul}, a method to identify excess energy in the data surrounding a signal assuming a Gaussian noise background, etc. If a glitch is present, a human expert will manually identify the time and frequency region of the glitch for downstream data analyses such as glitch removal or noise subtraction~\citep{Davis:2022ird}, e.g. via the BayesWave algorithm~\citep{Cornish_2015}. 
Accurate identification of glitches and their time-frequency region is essential to ensure reliable estimates of \ac{GW} source properties. %BayesWave [28–30] is algorithms for noise mitigation and subtraction that have been used to model and subtract glitches for LVK analyses [1–4] since the second observing run (O2) [22, 28] and is the most commonly used algorithm to date.
%If a glitch occurs near or overlaps with a \ac{GW} signal, the glitch may need to be removed before the source properties of the \ac{GW} signal can be reliably estimated~[Davis et al]. 
%A famous example is the loud glitch that occurred near the merger time of \ac{GW}170817, which needed to be removed before the source sky location estimate could be %reliably established that was essential for the associated electromagnetic campaign that resulted in the detection 
%of the associated kilonova, x-ray and radio emissions. This demonstrated the importance of noise mitigation and glitch removal for reliable parameter estimations.
% human interpretation can
%\hf{It might be worth mentioning (either in this paragraph or the end of the previous paragraph) that in addition to the existing data quality tools, detection pipelines themselves (GstLAL, PyCBC, etc.) also perform their own data conditioning and event ranking to mitigate false positive triggers.}

However, human validation of a candidate, including the identification of the time-frequency region of any glitches present is time intensive and can introduce inconsistencies, making validation results difficult to reproduce.
%glitch modeling often requires human inputs on priors such as the frequency range and the time window to reduce the parameter space that needs to be analyzed.
An expected increase in event rate for future observing runs that will leverage more sensitive \ac{GW} detectors\footnote{https://observing.docs.ligo.org/plan/} will significantly exacerbate this challenge.
%and using an additional time series that witnesses the source of the glitch to subtract the relevant excess noise.
%Glitch mitigation or noise subtraction can be done by modeling glitches using the strain data and subsequently subtracting the modeled glitches~[Ref].
%BayesWave [28–30] is algorithms for noise mitigation and subtraction that have been used to model and subtract glitches for LVK analyses [1–4] since the second observing run (O2) [22, 28] and is the most commonly used algorithm to date.
%\MC{BayesWave models glitches based on only the gravitational-wave strain data using wavelets. It has been used to subtract glitches in LVK analyses since the second observing run (O2) [22, 28] and is the most commonly used algorithm to date. Conversely, gwsubtract was used to subtract glitches in an LVK analysis for the first time in O3 [3]. However, this method was previously used for broadband noise subtraction in O2 [40].} 

In recent years, machine learning techniques and their applications have shown promising potential in many scientific fields including \ac{GW} astronomy. 
In particular, a machine learning technique for image segmentation known as \ac{YOLO}~\citep{redmon2016lookonceunifiedrealtime}, a convolutional neural network based technique, has been gaining popularity due to its efficiency and accuracy in locating objects of interest in the form of pixel coordinates from input images~\citep{wang2024yolov1}. 

In this work, we present as a proof of concept a novel algorithm based on \ac{YOLO} for the automatic identification, classification, and localization of glitches in time and frequency. The algorithm, referred to as \texttt{GSpyNetTreeS}, is an extension to the \texttt{GSpyNetTree} algorithm. In addition to classifying multiple glitches in \ac{GW} strain data simultaneously, \texttt{GSpyNetTreeS} is capable of estimating the frequency ranges and time windows during which the glitches occur. The estimates of the frequency and time ranges can then be used directly as inputs for downstream tasks such as noise subtraction. Incorporating \texttt{GSpyNetTreeS} in candidate vetting will enable fast, automatic and fully reproducible identification of the presence of data quality issues and reduce the need for human expert input, contributing to a more robust work flow. 

This manuscript is organized as follows: in Section~\ref{sec:gssec}, we will describe the development of \texttt{GSpyNetTreeS} including the generation of samples for training and testing; in Section~\ref{sec:performance}, we will present the performance of \texttt{GSpyNetTreeS} on the testing samples; in Section ~\ref{sec:conclusion} we conclude with a discussion of potential future applications.
\section{\bf \texttt{GS\lowercase{py}N\lowercase{et}T\lowercase{ree}S}}\label{sec:gssec}
Deep learning algorithms such as convolutional neural networks have had numerous prior applications to \ac{GW} astronomy, including real-time validation of events~\citep{Cabero:2020eik, Chan_2024, Abbott_2022, Raza:2023gyv}, the identification of detector glitches~\citep{Alvarez-Lopez:2023dmv, Zevin:2016qwy, Zevin:2023rmt, glanzer_2021_5649212, Wu_2025, Koyama_2024}, the detection of astrophysical signals~\citep{PhysRevLett.120.141103, Chan:2019fuz, Skliris:2020qax, Mishra:2022ott}, \ac{GW} source property estimation~\citep{Gabbard:2019rde, Langendorff:2022fzq, Green:2020hst}, sky localization~\citep{Chatterjee:2022ggk},  signal denoising~\citep{PhysRevD.108.043024}, and detector noise transient generation~\citep{Powell:2022pcg} (for a review, see~\cite{Cuoco:2020ogp}). Compared to more traditional methods for \ac{GW} data analysis, an advantage of deep learning algorithms is that they are relatively computationally cheap to run while the performance is comparable~\citep{heaton2018ian}.

\texttt{GSpyNetTreeS} leverages the \ac{YOLO} family, a class of deep learning algorithms designed for object detection in images with convolutional neural networks. Object detection in this context means the identification, the localization and the classification of objects of interest in a given image. \ac{YOLO} addresses these issues as a single regression problem, predicting both the locations and class probabilities of objects in input images. This approach allows for fast, efficient and accurate detection, making it suitable for real-time applications. 
%Compared to other existing object detection algorithms such as X and Y, \ac{YOLO} has the advantages of being low in computation cost while performance is good. \ac{YOLO} as a solution for object detection is known for its speed and accuracy.
%\MC{We reframe object detection as a single regression problem, straight from image pixels to bounding box coordinates and class probabilities. Using our system, you only look once (YOLO) at an image to predict what objects are present and where they are.}

Many versions of \ac{YOLO} have been developed in the literature, from \ac{YOLO} to \ac{YOLO}v11~\citep{khanam2024yolov11overviewkeyarchitectural} and beyond~\citep{cheng2024yoloworldrealtimeopenvocabularyobject}. Each version incorporates innovations to address different issues and challenges (see~\cite{Terven_2023, JIANG20221066, alif2024yolov1, wang2024yolov1} for reviews and~\citep{Soni:2025fqu} for a complementary application of \ac{YOLO} for \ac{GW} detection). In this work, we adopt an architecture that is closest to \ac{YOLO}v3 due to its balance of speed, accuracy and ease of implementation~\citep{redmon2018yolov3incrementalimprovement} while specific changes are made to allow \texttt{GSpyNetTreeS} to detect, locate and classify glitches and \acp{GW} given input data. 

\begin{figure}
\gridline{
    \fig{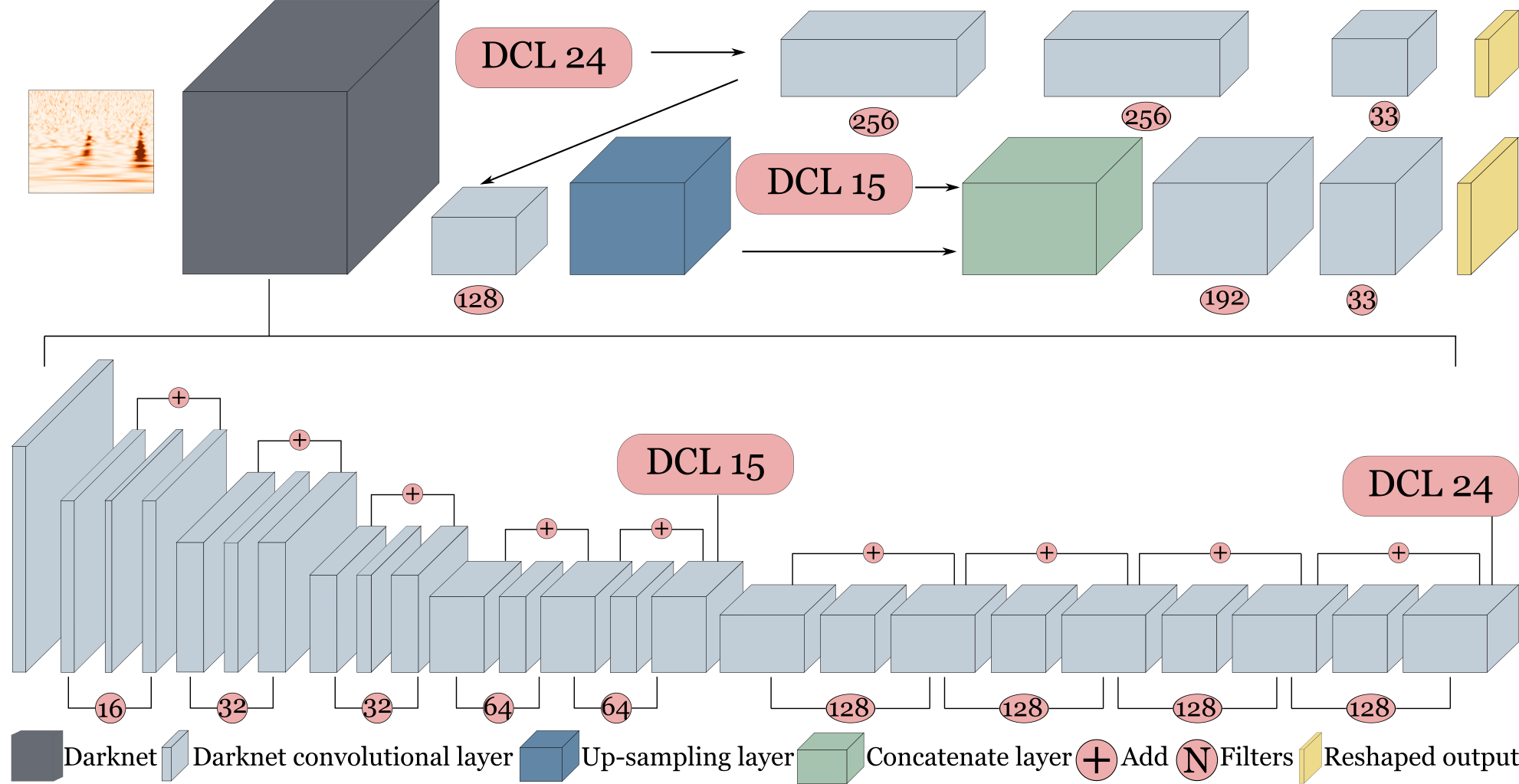}{0.45\textwidth}{(a)\label{fig:overallarchi}}
    }
\gridline{
    \fig{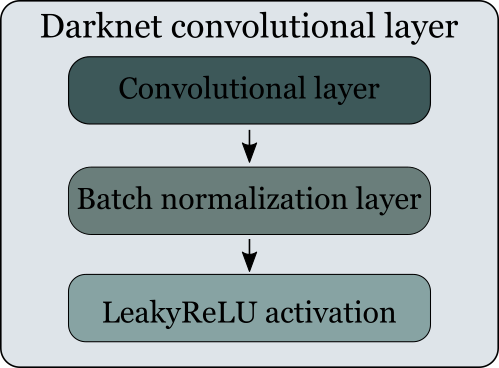}{0.28\textwidth}{(b)\label{fig:darknet}}
   }
\caption{The architecture of \texttt{GSpyNetTreeS}. The input to \texttt{GSpyNetTreeS} is a time-frequency representation image of \ac{GW} data. \texttt{GSpyNetTreeS} consists of two major components. The first component is a simplified Darknet with 24 Darknet convolutional layers. The Darknet generates two sets of outputs from its 15th and 24th layer. The outputs will be further processed by the second component of \texttt{GSpyNetTreeS}, which consists of a few Darknet convolutional layers, an up-sampling layer and a concatenate layer. The outputs are then passed onto two output layers, each of which is specialized for regions of excess power of different sizes (see Section~\ref{sec:training}). The lower panel shows the layers used to construct a Darknet convolutional layer (DCL).}
\label{fig:archi}
\end{figure}

As shown in Figure~\ref{fig:archi}, \texttt{GSpyNetTreeS}'s architecture includes a Darknet component for processing time-frequency representation of \ac{GW} detector data (see Section~\ref{sec:training}). The outputs from the Darknet component are then passed on to subsequent layers for further processing. The outputs of \texttt{GSpyNetTreeS} are two multi-dimensional arrays, containing information on the frequency boundaries, time windows and classification for glitches and \acp{GW} detected in input images, allowing \texttt{GSpyNetTreeS} to automatically provide useful information for event validation and downstream data analysis tasks such as glitch removal or noise subtraction. An example is shown in Figure~\ref{fig:groundtruth_blf_example}.
%\begin{figure}
%    \centering
%    \subfigure[]{%
%        \includegraphics[width=\columnwidth]{Plot/obj_archi.png}
%        \label{fig:overallarchi}
%    }\qquad
%    \subfigure[]{%   
%        \includegraphics[width=0.6\columnwidth]{Plot/dcc_archi_.png}
%        \label{fig:darknet}
%   }%\qquad
   %\subfloat[]{%   
   %     \includegraphics[width=\columnwidth]{make_plots/paper_plots/ft_cm.pdf}
   %     \label{fig:cmcm2}
   %}
%\caption{The architecture of \texttt{GSpyNetTreeS}. The input to \texttt{GSpyNetTreeS} is a time-frequency representation image of \ac{GW} data. \texttt{GSpyNetTreeS} consists of two major components. The first component is a simplified Darknet with 24 Darknet convolutional layers. The Darknet generates two sets of outputs from its 15th and 24th layer. The outputs will be further processed by the second component of \texttt{GSpyNetTreeS}, which consists of a few Darknet convolutional layers, an up-sampling layer and a concatenate layer. The outputs are then passed onto two output layers, each of which is specialized for regions of excess power of different sizes (see Section~\ref{sec:training}). The lower panel shows the layers used to construct a Darknet convolutional layer (DCL). 
%\label{fig:archi}}
%\end{figure}

\subsection{Data preparation}\label{sec:training}
To train \texttt{GSpyNetTreeS} for the identification, localization and classification of glitches and \acp{GW}, we use the \texttt{GSpyNetTree} data set described in~\cite{Alvarez-Lopez:2023dmv}. This data set consists of glitches observed in the LIGO Hanford and LIGO Livingston detectors during \ac{O3}~\citep{LIGOScientific:2020ibl, KAGRA:2021vkt}, obtained from the Gravity Spy classifications~\citep{glanzer_2021_5649212} via LIGO-DV web~\citep{Areeda:2016mee}. As a proof of concept we include in our data sets only Blips, Low frequency blips, Low frequency lines, Scattering and Scratchy\footnote{Note that LIGO detector glitch names are often derived from their time-frequency morphology (e.g. Blips, Lines) or source (e.g. Scattering)} as these are some of the most commonly occurring glitch classes~\citep{LIGO:2021ppb, LIGO:2024kkz}. 
The \ac{GW} signals were simulated assuming the waveform model IMRPhenomPv2~\citep{PhysRevD.93.044007, PhysRevD.93.044006} and added to segments of 64-second quiet detector data from both the LIGO Hanford and LIGO Livingston detectors during previous observing runs, using the inspiral injection module of LALSuite~\citep{lalsuite}.
For this proof-of-principle study, to compare performance against commonly occurring glitches we focus on higher mass \ac{BBH} sources which are of $\lesssim$ 1-second duration in the detectors' sensitive frequency range. \ac{GW} examples were uniformly drawn from a total merger mass range of 5M to 350M, with individual masses ranging from 2M$_\odot$ to 175M$_\odot$, an SNR range of 8 to 35, and individual component spins ranging from 0.05 to 0.95.
%\hf{The description of the GW injection parameters are currently commented out, but it would informative to include it in the text, particularly to highlight that the injections focus on to BBH and IMBH, signals .}

\begin{figure}
\gridline{
    \fig{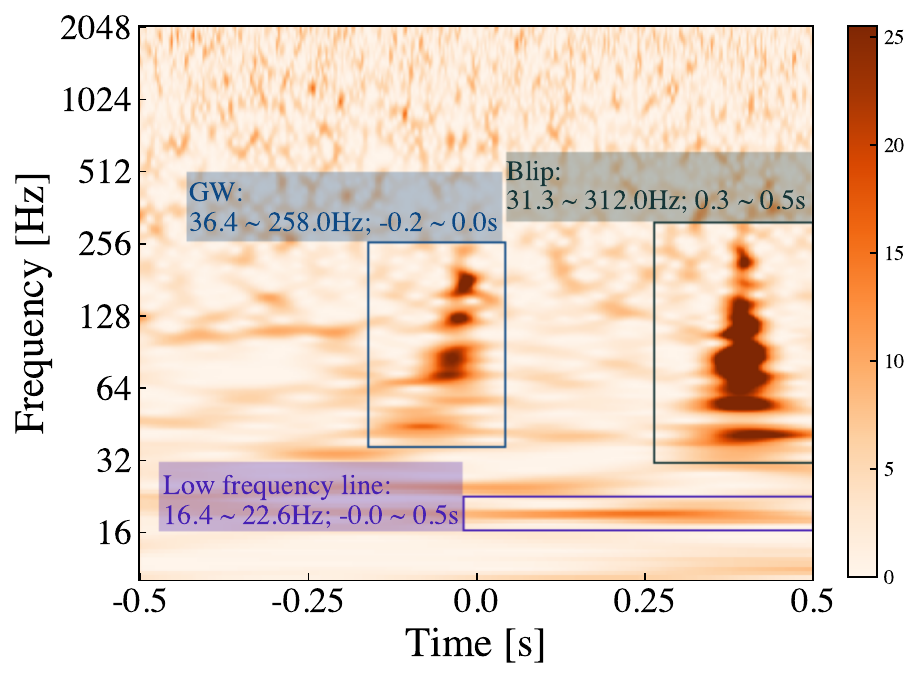}{0.45\textwidth}{(a)\label{fig:groundtruth_sample}}
    }
\gridline{
    \fig{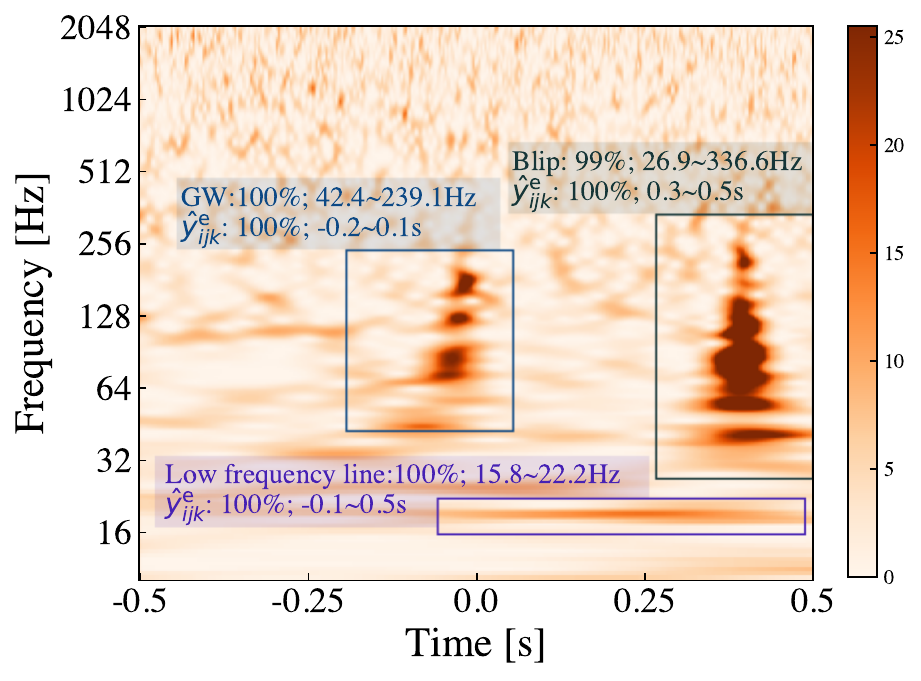}{0.45\textwidth}{(b)\label{fig:testing_sample}}
   }
\caption{Images of time-frequency representation of a time series from LIGO Hanford. The images show three regions of excess power that meet the criteria defined in  Section~\ref{sec:training}. A simulated \ac{GW} signal 
%\hf{I found it a bit confusing to refer to this signal as being in the middle (on my first read, I assumed the text was referring to the low frequency line glitch) - perhaps referring them by the bounding box/text colour would be clearer} 
from a binary black hole merger with component masses 35.6 and 40.3\,$\mathrm{M}_{\odot}$ is shown in the blue time-frequency box. The other two time-frequency boxes highlight real glitches observed by the detector: a Blip (green box) and Low frequency line (red box). The upper panel shows the `ground truth': manually drawn bounding boxes, with human expert judgment used to assess the most appropriate time-frequency bounds. The lower panel shows the corresponding bounding boxes predicted by the algorithm, including predicted frequency ranges and time windows. $\hat{y}^{\mathrm{e}}_{ijk}$ indicates the confidence of \texttt{GSpyNetTreeS} that the predicted bounding boxes surround a region of excess power (see Section~\ref{sec:loss}). 
%\hf{I think that the bottom panel speaks for itself, and the top panel can be omitted. I would also increase the font size in the shaded boxes as it's a little hard to read (suggestion: same font size as the (a) and (b) subtitles).}
\label{fig:groundtruth_blf_example}}
\end{figure}

Time-frequency representations of the data set are then generated using the Q-transform~\citep{Chatterji:2004qg}, a time-frequency visualization tool commonly used for characterizing \ac{GW} strain data. The Q-transform is a modification of the short-time Fourier transform that tiles the time-frequency plane with pixels of constant quality factor Q. This is achieved by using analysis windows with durations inversely proportional to the frequency, resulting in different resolutions at different frequencies. In this proof-of-concept work, we use a time-frequency representation of 1 second in duration with a quality factor Q of 25, consistent the feature sets of image classification algorithms \texttt{Gravity Spy}~\citep{Zevin:2016qwy, Zevin:2023rmt, Wu_2025} and \texttt{GSpyNetTree}~\citep{Alvarez-Lopez:2023dmv}, which leverage similar images of different durations.
The time-frequency representations are converted to images of 280 pixels by 340 pixels, which are then normalized such that the values of the pixels range from 0 to 25.5, with higher values indicating larger energy as illustrated in Figure~\ref{fig:groundtruth_blf_example}.

\begin{figure}
\includegraphics[width=0.45\textwidth]{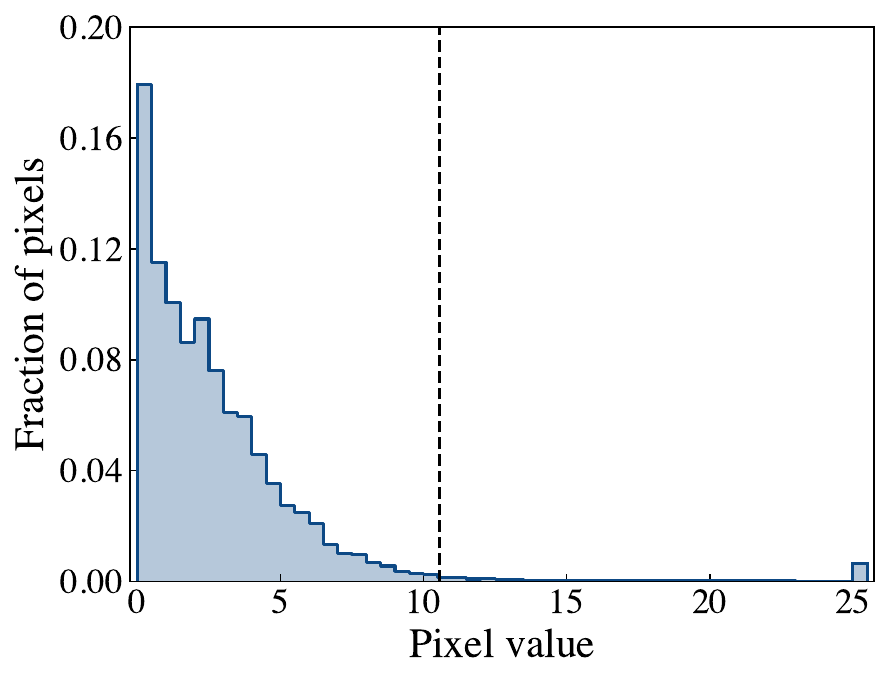}
\caption{A histogram showing the distribution of pixel values for all images of time-frequency representation of \ac{GW} detector data in the training samples. The vertical line indicates the threshold for loud pixels (i.e., 10.5), which is greater than $98\%$ of all pixels.
\label{fig:clusterthreshold}}
\end{figure}

%It is possible that using omega scans of 4 lengths may provide the model with more comprehensive information on the types of glitches present in the data. However, the primary focus of GSpyNetTreeS is to identify and locate all regions of excess power given some gravitational detector data.
In order to train \texttt{GSpyNetTreeS}, we first need to define the time-frequency region corresponding to glitches and \acp{GW}.
We considered glitches and \acp{GW} loosely as regions of excess power of different morphologies present in time-frequency representation of \ac{GW} detector data. We defined a region of excess power in the time-frequency representation as a cluster of image pixels within which the number of connected neighboring loud pixels is $\geq 28$. Loud pixels are defined as image pixels where the normalized pixel energy is $\geq 10.5$. This value is chosen such that loud pixels are image pixels with energy greater than $98\%$ of all pixels in the data set as shown in Figure~\ref{fig:clusterthreshold}. Based on these criteria, however, some simulated \ac{GW} signals were too weak and are thus not recognized as regions of excess power. These signals are therefore excluded from our data. Excluding these signals will undermine \texttt{GSpyNetTreeS}'s capability to detect weak \acp{GW}, however, this is not a limitation to \texttt{GSpyNetTreeS}'s intended purpose for the identification of data quality issues for event validation. In addition, these threshold choices 
%determine only the sensitivity of the algorithm to glitches and \acp{GW} and 
can be easily adjusted prior to training to incorporate weaker examples, if desired.

%Each of the time-frequency representation images is then labeled manually. 
The time-frequency region containing each instance of excess power identified by the above criteria was then labeled manually as a bounding box, with human expert judgment used to assess the most appropriate time-frequency bounds. 
The class (e.g. glitch type) was labeled according to the \texttt{GSpyNetTree} data set classification. Figure~\ref{fig:groundtruth_sample} shows an example generated from \ac{O3} LIGO Hanford data, with three manually identified and labeled regions of excess power.  
In total, 13\ 000 unique time-frequency representation images were generated. The time-frequency representation images were then augmented by applying a random time shift of at most 0.25\,s to the \ac{GW} strain data in time prior to generation of time-frequency representation of the data, resulting in 23\ 749 time-frequency representation images in total. This augmentation improves the algorithm's robustness to changes in noise background, as demonstrated in~\citep{Alvarez-Lopez:2023dmv}. The number of regions of excess power identified for each class (\acp{GW} and glitches) is given in Table~\ref{table:roepn}. Approximately $10\%$ of the data were randomly selected and reserved for testing, and an additional $10\%$ were reserved for validation during training. 
%\hf{I'm confused about the usage of the word `manual' to describe the bounding box labelling. Were the bounding boxes decided by human visual inspection (which would be my first assumption) or was it automated via an algorithm? How were the boundaries decided, and for how many representation images were manual bounding boxes drawn?}

\begin{table}[]
\centering
\begin{tabular}{lcccc}
\toprule
Morphology         & Training & Validation & Testing \\
\hline
\ac{GW}            & 11479    & 1249       & 1385    \\
Blip               & 3993     & 464        & 513     \\
Low frequency blip & 3440     & 374        & 449     \\
Low frequency line & 5767     & 651        & 667     \\
Scattering         & 1832     & 198        & 250     \\
Scratchy           & 19382    & 2124       & 2419    \\
Images             & 19236    & 2138       & 2375     \\
\hline
\end{tabular}
\caption{The number of regions of excess power identified for each class based on morphology.
The last row indicates the number of images for training, validation and test
To train \texttt{GSpyNetTreeS}, approximately 10\% of the total samples are randomly drawn separately for validation and testing. \label{table:roepn}}
\end{table}

\begin{figure}
\includegraphics[width=0.45\textwidth]{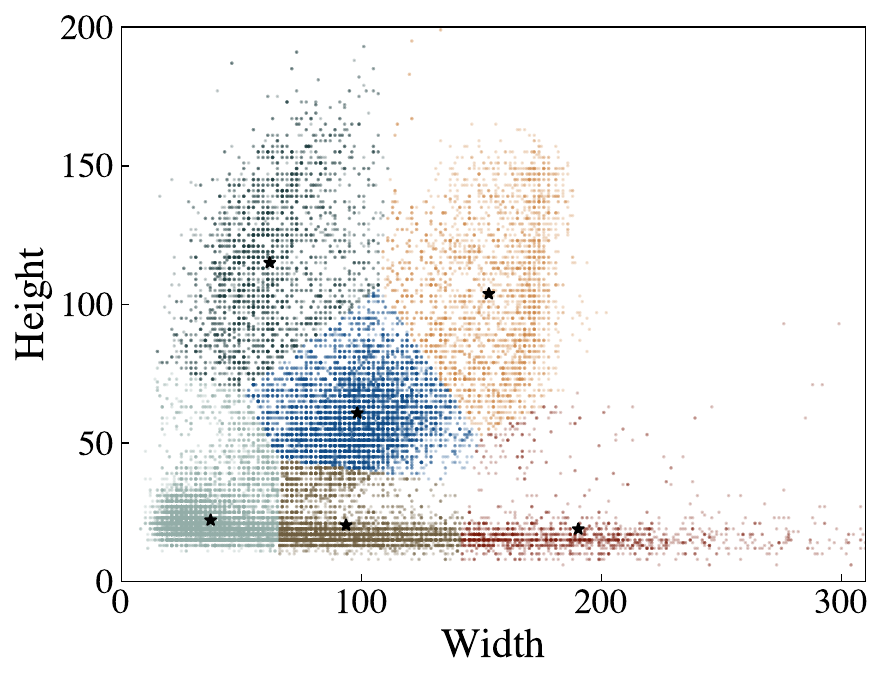}
\caption{A scatter plot showing the distribution of the widths and heights in pixels for all the bounding boxes in the images for training. K-means clustering is applied to group bounding boxes into different clusters, with cluster members shown in different colours. The centroids of the clusters are then taken as the \textit{typical size} of anchor boxes, which are used as reference boxes for \texttt{GSpyNetTreeS}. 
%\hf{What do the different colours mean? The colours are also similarly muted and can be difficult to differentiate; I suggest varying the contrast and brightness to make the colours clearer.}
\label{fig:kmeans}}
\end{figure}

Image segmentation algorithms require clear coaching on their target output, which in our case is time-frequency bounding boxes. In order to help simplify \texttt{GSpyNetTreeS}'s task we fed it the typical sizes of time-frequency boxes by applying K-means clustering to bounding boxes' widths and heights from the images set aside for training. We identified six distinct groups of time-freqeuncy bounding boxes, as illustrated in Figure~\ref{fig:kmeans}. The centroids of these bounding box clusters form \textit{anchor boxes}, which act as reference bounding boxes for each cluster. 

For each time-frequency representation image, six grids covering the entire image are then generated, with each associated with one of the six anchor boxes.
The bounding box for each manually identified region of excess power is then matched to a grid cell from a grid associated with the anchor box that most closely matches the bounding box's dimensions, as determined using K-means clustering. The specific grid cell responsible for the region of excess power is determined by locating the grid cell containing the center of that region. Each grid cell can be associated with one region of excess power at most. 

The assigned grid cell is then labeled with the bounding box’s normalized width, height, and center coordinates relative to the cell, as well as the predicted morphological class of the region. The use of anchor boxes allows each grid to specialize in detecting objects of a specific size range, thus improving the model's performance in identifying differing regions of excess power in time-frequency images. To further optimize the performance of \texttt{GSpyNetTreeS} for regions of excess power of different sizes, these grids are divided into two groups with different number of grid cells. The bounding box information in each group of the grids is then the target output for one of the output layers as shown in Figure~\ref{fig:archi}. The number of grid cells for each grid and the size of the associated anchor boxes are given in Table~\ref{table:gridsize}. 

\begin{table}
\centering
\begin{tabular}{llll}
\toprule
Output                 &    Grid                   & Grid cells                           & Anchor box               \\         
\hline
                       &    1                      &                                      & (61.67, 114.97)               \\
          1            &    2                      &(17,  21)                             & (152.95, 103.72)              \\
                       &    3                      &                                      & (190.53, 18.91)               \\

                       &    4                      &                                      & (37.17,  22.15)               \\
          2            &    5                      &(35,  42)                             & (93.65,  20.28)               \\
                       &    6                      &                                      & (98.22,  60.72)               \\
\hline
\end{tabular} 
\caption{The number of grids, grid cells and the sizes of the associated anchor boxes. The first column
indicates the outputs from \texttt{GSpyNetTreeS} (see Figure~\ref{fig:archi}). Each output from \texttt{GSpyNetTreeS} contains predicted bounding box information for three grids. The second column indicates the grid and the third column the number of grid cells, i.e., (row, column) for each of the grids.  The sizes of the associated anchor boxes in pixel are shown in the fourth column. \label{table:gridsize}}
\end{table}

An example prediction from \texttt{GSpyNetTreeS} is given in Figure~\ref{fig:testing_sample}. 
%\hf{Perhaps as a replacement to the top panel of Figure 2 or an additional subplot to Figure 4, it would be helpful to have a plot illustrating the grid cell method to supplement the text.}

\subsection{Loss function}\label{sec:loss}
Machine learning algorithms are generally trained by minimizing their loss functions. Loss functions measure and quantify the error between a machine learning algorithm's outputs given an input and the ground truth. During training, small subsets of training data are passed through the algorithm in multiple iterations. The back-propagation algorithm then calculates the gradient of the loss function with respect to the trainable parameters of the algorithm, such as the weights and biases of the neurons in each layer. These parameters are then updated using a gradient descent algorithm in order to iteratively minimize the loss-function. Once this process converges and the loss function is minimized,  the training is considered complete. The algorithm will be able to take in input in the form of new data and generate outputs that best correspond to the learned mappings between input and output during training. An appropriate loss function is therefore important to achieve the desired performance. 

\texttt{GSpyNetTreeS} is designed to detect, locate and classify regions of excess power in a given time-frequency representation image representing \ac{GW} detector data. This involves 1. identifying the existence of regions of excess power, 2. determining the positions of the corresponding bounding boxes, 3. estimating the width and height of the boxes, and 4. classifying the morphology of the detected regions of excess power. 
%\hf{This is a very helpful list that makes it easy to understand GSpyNetTreeS's objectives and helps guide the reader through the subsequent sections. Could a version of this list be added earlier in the paper, perhaps in Section 2 where GSpyNetTreeS is introduced?} 
As such, the loss function must incorporate each of these objectives. 
For each input time-frequency representation image, we therefore employ the following loss function,
\begin{equation}\label{eq:loss}
\begin{aligned}
L_{\mathrm{Total}}&(\textbf{y}^{\mathrm{e}}, \textbf{y}^{\mathrm{p}}, \textbf{y}^{\mathrm{wh}}, \textbf{y}^{\mathrm{C}}, \hat{\textbf{y}}^{\mathrm{e}}, \hat{\textbf{y}}^{\mathrm{p}}, \hat{\textbf{y}}^{\mathrm{wh}}, \hat{\textbf{y}}^{\mathrm{C}})\\ &=  L_{\mathrm{e}}(\textbf{y}^{\mathrm{e}}, \hat{\textbf{y}}^{\mathrm{e}}) + L_{\mathrm{p}}(\textbf{y}^{\mathrm{p}}, \hat{\textbf{y}}^{\mathrm{p}})+\\ &\hspace{1.3em}L_{\mathrm{wh}}(\textbf{y}^{\mathrm{wh}}, \hat{\textbf{y}}^{\mathrm{wh}})+ L_{\mathrm{C}}(\textbf{y}^{\mathrm{C}}, \hat{\textbf{y}}^{\mathrm{C}}),
\end{aligned}
\end{equation}
where $L_{\mathrm{Total}}$ is the total loss and $L_{\mathrm{e}}$, $L_{\mathrm{p}}$, $L_{\mathrm{wh}}$ and $L_{\mathrm{C}}$ are the losses for the absence or presence of any regions of excess power in the image, the positions of the corresponding bounding boxes, the widths and heights of the boxes and the classification of the regions of excess power respectively. The losses are defined below, 
\begin{equation}\label{eq:loss2}
\begin{aligned}
&L_{\mathrm{e}}(\textbf{y}^{\mathrm{e}}, \hat{\textbf{y}}^{\mathrm{e}}) = \frac{1}{N_{\mathrm{g}}}\sum^{N_{\mathrm{g}}}_{k}\frac{1}{N_{\mathrm{gc}}(k)}\times\\
&\hspace{1.0em} \sum^{N_{\mathrm{gc}}(k)}_{i,j}\{m_{ijk} [y^{\mathrm{e}}_{ijk} \log{\hat{y}^{\mathrm{e}}_{ijk}} + (1 - y^{\mathrm{e}}_{ijk})\log(1-\hat{y}^{\mathrm{e}}_{ijk})]
+\\
&(1-m_{ijk})n_{ijk}[y^{\mathrm{e}}_{ijk} \log{\hat{y}^{\mathrm{e}}_{ijk}} + (1 - y^{\mathrm{e}}_{ijk})\log(1-\hat{y}^{\mathrm{e}}_{ijk})]
\};\\
&L_{\mathrm{p}}(\textbf{y}^{\mathrm{p}}, \hat{\textbf{y}}^{\mathrm{p}}) = \frac{1}{N_{\mathrm{g}}}\sum^{N_{\mathrm{g}}}_{k}\frac{1}{N_{\mathrm{gc}}(k)}\times \\
&\hspace{1.0em} \sum^{N_{\mathrm{gc}}(k)}_{i,j}m_{ijk} \{ (y^{\mathrm{px}}_{ijk} - \hat{y}^{\mathrm{px}}_{ijk})^2 + (y^{\mathrm{py}}_{ijk} - \hat{y}^{\mathrm{py}}_{ijk})^2 \};\\
&L_{\mathrm{wh}}(\textbf{y}^{\mathrm{wh}}, \hat{\textbf{y}}^{\mathrm{wh}}) = \frac{1}{N_{\mathrm{g}}}\sum^{N_{\mathrm{g}}}_{k}\frac{1}{N_{\mathrm{gc}}(k)}\times\\
&\hspace{1.0em} \sum^{N_{\mathrm{gc}}(k)}_{i,j}m_{ijk} \{(y^{\mathrm{w}}_{ijk} - \hat{y}^{\mathrm{w}}_{ijk})^2 + (y^{\mathrm{h}}_{ijk} - \hat{y}^{\mathrm{h}}_{ijk})^2\};\\
&L_{\mathrm{C}}(\textbf{y}^{\mathrm{C}}, \hat{\textbf{y}}^{\mathrm{C}}) = \frac{1}{N_{\mathrm{g}}}\sum^{N_{\mathrm{g}}}_{k}\frac{1}{N_{\mathrm{gc}}(k)}\times \\
&\hspace{1.0em} \sum^{N_{\mathrm{gc}}(k)}_{i,j}m_{ijk} \sum^{N_\mathrm{class}}_{c} y^{\mathrm{C}}_{ijkc} \log({\hat{y}^{\mathrm{C}}}_{ijkc}).
\end{aligned}
\end{equation}
In the above equations, $\textbf{y}^{\mathrm{e}}$ is a three dimensional array representing the ground truth for the presence or absence of regions of excess power in the entire image. Its element $y^{\mathrm{e}}_{ijk}$ represents the ground truth for the presence or absence of any region of excess power associated with the grid cell at the $i$\,th row and $j$\,th column of the $k$\,th grid. The value of $y^{\mathrm{e}}_{ijk}$ can be either 1 (presence) or 0 (absence). $\textbf{y}^{\mathrm{p}}$ is an array composed of two separate three-dimensional arrays, $\textbf{y}^{\mathrm{px}}$ and $\textbf{y}^{\mathrm{py}}$ representing the ground truth coordinates of the center of the bounding boxes corresponding to the regions of excess power. For a bounding box of which the center falls within the grid cell at the $i$\,th row and $j$\,th column of the $k$\,th grid, 
%If there is a region of excess power in a grid cell at the $i$th row and $j$th column of the $k$th grid, $y^{\mathrm{l}}_{ijk}$ indicates the ground truth position of the center of the corresponding bounding box relative to the grid cell.
both the horizontal coordinate $y^{\mathrm{px}}_{ijk}$ and vertical coordinate $y^{\mathrm{py}}_{ijk}$ are linearly continuous and are determined by the relative position of the center to the grid cell. A coordinate of (1, 1) for ($y^{\mathrm{px}}_{ijk}$, $y^{\mathrm{py}}_{ijk}$) indicates that the center is at the bottom rightmost of the grid cell while (0, 0) at the top leftmost. The widths and heights of all the bounding boxes are expressed as $\textbf{y}^{\mathrm{wh}}$, which is similarly composed of two separate three-dimensional arrays $\textbf{y}^{\mathrm{w}}$ and $\textbf{y}^{\mathrm{h}}$ indicating the widths and heights respectively. The width $y^{\mathrm{w}}_{ijk}$ and the height $y^{\mathrm{h}}_{ijk}$ of a bounding box associated with the grid cell at the $i$\,th row and $j$\,th column of the $k$\,th grid are expressed in relation to the anchor box associated with the grid, as given below
\begin{equation}\label{fig:widthheight}
\begin{aligned}
\mathrm{B_w} &= \mathrm{A_w}_k \mathrm{exp}(y^{\mathrm{w}}_{ijk});\\
\mathrm{B_h} &= \mathrm{A_h}_k \mathrm{exp}(y^{\mathrm{h}}_{ijk}),
\end{aligned}
\end{equation}
where $\mathrm{B_w}$, $\mathrm{B_h}$, $\mathrm{A_w}_k$ and $\mathrm{A_h}_k$ are the widths and heights for the bounding box and the associated anchor box respectively.

Finally, $\textbf{y}^{\mathrm{C}}$ is a multi-dimensional array. Its element $\textbf{y}^{\mathrm{C}}_{ijk}$ is a one-dimensional array where the number of elements is equal to the number of considered classes, corresponding to the ground truth classification of a region of excess power whose center falls within the grid cell at the $i$\,th row and $j$\,th column of the $k$\,th grid.
The corresponding outputs from \texttt{GSpyNetTreeS} for the grid cell at the $i$\,th row and $j$\,th column of the $k$\,th grid are $\hat{y}^{\mathrm{e}}_{ijk}$,  $\hat{y}^{\mathrm{px}}_{ijk}$, $\hat{y}^{\mathrm{py}}_{ijk}$,  $\hat{y}^{\mathrm{w}}_{ijk}$,  $\hat{y}^{\mathrm{h}}_{ijk}$  and  $\hat{\textbf{y}}^{\mathrm{C}}_{ijk}$. Unlike its ground truth counterpart $y^{\mathrm{e}}_{ijk}$, however, $\hat{y}^{\mathrm{e}}_{ijk}$ is a value ranging from 0 to 1 indicating the \texttt{GSpyNetTreeS}'s confidence that a region of excess power is present and associated with the grid cell at the $i$\,th row and $j$\,th column of the $k$\,th grid. In addition, similar to $\textbf{y}^{\mathrm{C}}_{ijk}$, $\hat{\textbf{y}}^{\mathrm{C}}_{ijk}$ is a one-dimensional array where each element ranges from 0 to 1 corresponding to \texttt{GSpyNetTreeS}'s confidence that the detected region of excess power belongs to each of the classes. The elements of $\hat{\textbf{y}}^{\mathrm{C}}_{ijk}$ sum to 1. In the equations, the summations sum over the number of grids $N_{\mathrm{g}}$, the number of grid cells in the $k$\,th grid $N_{\mathrm{gc}}(k)$, and the number of classes $N_{\mathrm{class}}$.
To allow for an optimal performance, the losses are modeled differently. In particular, we employ binary cross-entropy for $L_{\mathrm{e}}$, mean squared errors for both $L_{\mathrm{p}}$ and $L_{\mathrm{wh}}$, and categorical cross-entropy 
for $L_{\mathrm{c}}$.

For any given image, the number of grid cells that are not associated with any regions of excess power in the image will likely vastly outnumber regions of excess power. As a result, the loss function may disproportionately emphasize these empty grid cells, driving \texttt{GSpyNetTreeS} to overwhelmingly predict the absence of region of excess power. To account for this, a mask is applied to the loss function. If the grid cell at the $i$\,th row and $j$\,th column of the $k$\,th grid is associated with a region of excess power, the value of $m_{ijk}$ will be 1 and 0 otherwise. Applying the mask removes the effects of empty cells and allows the model to focus on grids cells associated with regions of excess power. Unfortunately, for $L_{\mathrm{e}}$, this may have the negative effect of achieving an algorithm that will always generate $\hat{y}^{\mathrm{e}}_{ijk}$ close to 1, over predicting the existence of regions of excess power. This is because for empty grid cells (i.e., $m_{ijk}$ equals 0), multiplying $m_{ijk}$ prevents the loss function from penalizing the algorithm 
for generating non-zero $\hat{y}^{\mathrm{e}}_{ijk}$. To counter this effect, we apply a term $n_{ijk}$ to the calculation of $L_{\mathrm{e}}$. The value of $n_{ijk}$ is determined by computing the value of \ac{IOU} (see Section~\ref{sec:performance}) between the predicted bounding boxes at the $i$\,th row and $j$\,th column of the $k$\,th grid and each ground truth bounding box in the image. If the \ac{IOU} for at least one ground truth bounding box and the predicted bounding box is greater than 0.3, $n_{ijk}$ is 1 and 0 otherwise. The losses are then averaged over the number of grid cells $N_{\mathrm{gc}}(k)$ in the $k$\,th grid and the number of grids $N_{\mathrm{g}}$.  %These loss terms are multiplied by $m_{ijk}$ and then averaged over the number of grid cells $N_{gc}(k)$ in a grid and the number of grids $N_g$. $\textbf{y}^{\mathrm{e}}$,  $\textbf{y}^{\mathrm{l}}$, $\textbf{y}^{\mathrm{wh}}$ and $\textbf{y}^{\mathrm{c}}$ are the target outputs for the existences of regions of excess power, the locations and widths and heights of the corresponding bounding boxes and the classification of the morphology respectively. The actual outputs from GSpyNetTreeS for these terms are tilted.
The training of \texttt{GSpyNetTreeS} is complete when the losses in Eq.~\ref{eq:loss2} are minimized.
\section{Performance}\label{sec:performance}
\begin{figure}
\gridline{
    \fig{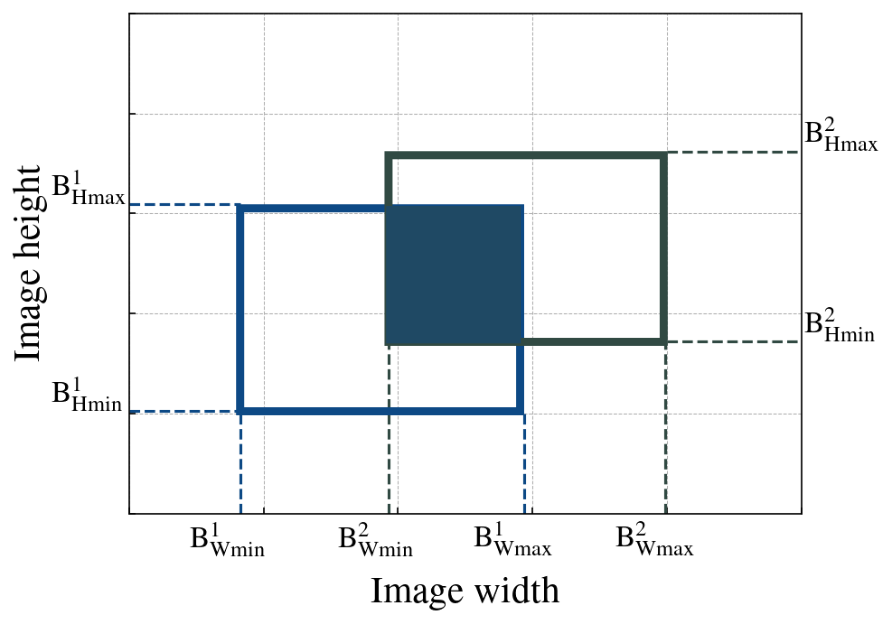}{0.45\textwidth}{(a)\label{fig:iou_overlap}}
    }
\gridline{
    \fig{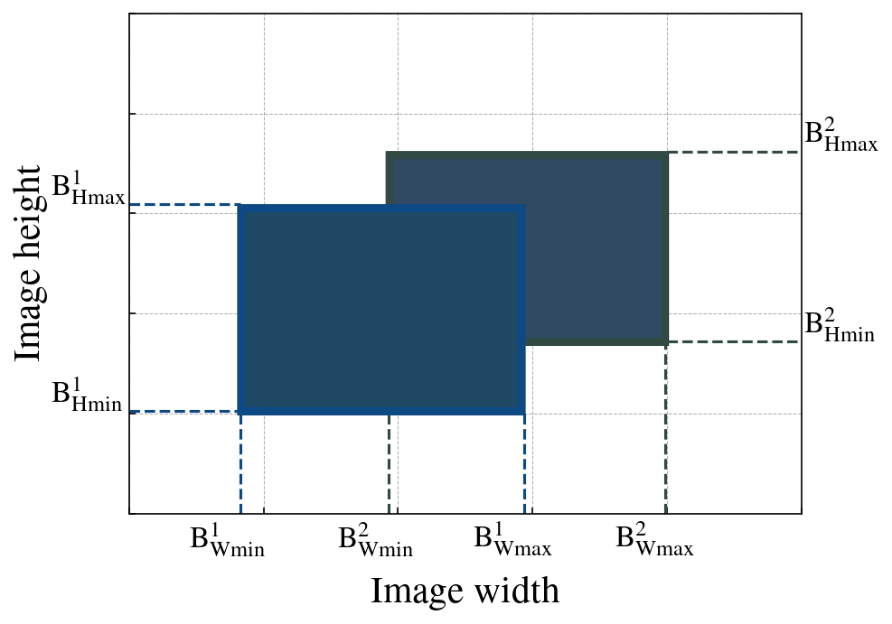}{0.45\textwidth}{(b)\label{fig:iou_union}}
   }
\caption{The definition of \ac{IOU}. Given two boxes $\mathrm{B}^{1}$ and $\mathrm{B}^{2}$, the shaded areas in the top and bottom panels show the overlapping area between the two boxes $\mathrm{A_{overlap}}$ and the combined area of the boxes $\mathrm{A_{union}}$ respectively.
\label{fig:iou}}
\end{figure}

To show the performance of \texttt{GSpyNetTreeS} on the testing samples given in Table~\ref{table:roepn}, it is necessary to introduce the concept of \ac{IOU}~\citep{rezatofighi2019generalizedintersectionunionmetric}. \ac{IOU} is a metric used to measure and evaluate the match between two boxes with given sizes and positions. For two boxes $\mathrm{B}^{1}$ and $\mathrm{B}^{2}$, \ac{IOU} is given by
\begin{equation}
\mathrm{IoU} = \frac{\mathrm{A_{overlap}}}{\mathrm{A_{union}}},\\
\end{equation}
where $\mathrm{A_{overlap}}$ and $\mathrm{A_{union}}$ are the overlapping area and the combined area of the boxes, as illustrated in Figure~\ref{fig:iou} and given by
\begin{equation}
\begin{aligned}
\mathrm{A_{overlap}} &=(\mathrm{B}^{1}_{\mathrm{W_{max}}} - \mathrm{B}^{2}_{\mathrm{W_{min}}})(\mathrm{B}^{1}_{\mathrm{H_{max}}} - \mathrm{B}^{2}_{\mathrm{H_{min}}})\\
\mathrm{A_{union}} &= (\mathrm{B}^{1}_{\mathrm{W_{max}}} - \mathrm{B}^{1}_{\mathrm{W_{min}}})(\mathrm{B}^{1}_{\mathrm{H_{max}}} - \mathrm{B}^{1}_{\mathrm{H_{min}}}) + \\
&\hspace{1.3em} (\mathrm{B}^{2}_{\mathrm{W_{max}}} - \mathrm{B}^{2}_{\mathrm{W_{min}}})(\mathrm{B}^{2}_{\mathrm{H_{max}}} - \mathrm{B}^{2}_{\mathrm{H_{min}}}) - \\
&\hspace{1.3em} \mathrm{A_{overlap}},
\end{aligned}
\end{equation} 
%Mathematically, they are given by
%\begin{equation}\label{eq:loss2}
%\begin{aligned}
%\mathrm{Area~of~Overlap} &= \max(0, \min(\mathrm{B^{1}_{Wmax}}, \mathrm{B^{2}_{Wmax}})) \times \\ \max(0, \min(\mathrm{B^{1}_{Hmax}}, \mathrm{B^{2}_{Hmax}}))
%\end{aligned}
%\end{equation}
where $\mathrm{B}^{1/2}_{\mathrm{W_{max/min}}}$ and $\mathrm{B}^{1/2}_{\mathrm{H_{max/min}}}$ are the coordinates of the boxes as shown in Figure~\ref{fig:iou}.
%\begin{figure}
%    \centering
%    \subfigure[]{%
%        \includegraphics[width=\columnwidth]{Plot/IOU_overlap.png}
%        \label{fig:iou_overlap}
%    }\qquad
%    \subfigure[]{%   
%        \includegraphics[width=\columnwidth]{Plot/IOU_union.png}
%        \label{fig:iou_union}
%   }%\qquad
   %\subfloat[]{%   
   %     \includegraphics[width=\columnwidth]{make_plots/paper_plots/ft_cm.pdf}
   %     \label{fig:cmcm2}
   %}
%\caption{The definition of \ac{IOU}. Given two boxes $\mathrm{B}^{1}$ and $\mathrm{B}^{2}$, the shaded areas in the top and bottom panels show the overlapping area between the two boxes $\mathrm{A_{overlap}}$ and the combined area of the boxes $\mathrm{A_{union}}$ respectively.
%\label{fig:iou}}
%\end{figure}
For any given pair of boxes, \ac{IOU} ranges from 0 to 1, with 0 indicating that the boxes do not have any overlap and 1 indicating that the boxes completely overlap with each other. 

For an input image, multiple nearby grid cells in the outputs from \texttt{GSpyNetTreeS} may pick up the same region of excess power, generating different bounding boxes, essentially containing the same information, albeit with varying degree of confidence for the presence of the region of excess power (i.e., $\hat{y}^{\mathrm{e}}_{ijk}$). To select the bounding boxes that represent the algorithm's most confident detections, a threshold $\mathrm{T_{e}}$ on detection confidence (i.e., $\hat{y}^{\mathrm{e}}_{ijk}$) is selected. The values of $\hat{y}^{\mathrm{e}}_{ijk}$ for all grid cells passing $\mathrm{T_{e}}$ are ranked and the bounding box with the maximum confidence is then marked as the first valid detection. The values of 
\ac{IOU} between the selected bounding box and each of the remaining bounding boxes will then be computed. If for a bounding box, the value of \ac{IOU} is larger than a pre-selected threshold $\mathrm{T_{IoU}}$, the bounding box will be considered repeated detection and be discarded. 
After all the bounding boxes with \ac{IOU} $>\mathrm{T_{IoU}}$ are discarded, the remaining bounding boxes will have their $\hat{y}^{\mathrm{e}}_{ijk}$ ranked again and the bounding boxes with the maximum $\hat{y}^{\mathrm{e}}_{ijk}$ will be marked as the second valid detection. This process will repeat until no \ac{IOU} between any pair of bounding boxes is larger than $\mathrm{T_{IoU}}$ and the maximum $\hat{y}^{\mathrm{e}}_{ijk}$ for the remaining bounding boxes is $<\mathrm{T_{e}}$. The selected set of bounding boxes then represents the algorithm's confident detections of regions of excess power in the input image. In the remainder of this paper, we use a $\mathrm{T_{e}}$ of 0.5 and a $\mathrm{T_{IoU}}$ of 0.3. 

\begin{figure}
\includegraphics[width=0.45\textwidth]{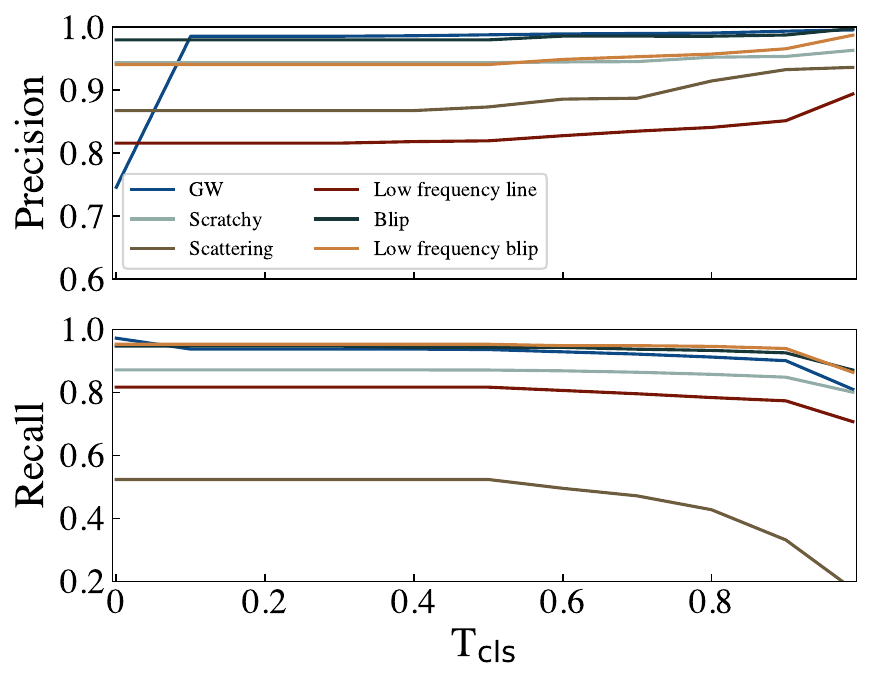}
\caption{The performance of \texttt{GSpyNetTreeS} shown as curves of precision (upper panel) and recall (lower panel). The curves are produced using a $\mathrm{T_{e}}$ of 0.5 and a $\mathrm{T_{IoU}}$ of 0.3 respectively. 
\label{fig:roc_alternative}}
\end{figure}

We will first show the performance of \texttt{GSpyNetTreeS} in detecting the presence of regions of excess power as well as the classifications of the detected regions of excess power. Since for \texttt{GSpyNetTreeS}, a correct identification of a region of excess power requires the correct classification and estimation of the position and size of the bounding box, we define true positives as valid detections with the following requirements: 1. the \ac{IOU} between the predicted bounding box and the ground truth bounding box is $\geq \mathrm{T_{IoU}}$, and 2. the morphology is correctly classified based on a threshold $\mathrm{T_{cls}}$. False positives are defined as valid detections with the following conditions: 1. the predicted bounding box either does not have an associated ground truth bounding box where the \ac{IOU} is $\geq \mathrm{T_{IoU}}$, or 2. the classification of the morphology is incorrect based on $\mathrm{T_{cls}}$. False negatives are defined as a ground truth region of excess power with the following conditions: 1. no corresponding predicted bounding box with an \ac{IOU} $\geq \mathrm{T_{IoU}}$, or 2. the classification of the morphology is incorrect based on $\mathrm{T_{cls}}$ for all predicted bounding boxes with an \ac{IOU} $\geq \mathrm{T_{IoU}}$. However, it should be noted that true negative does not have a straightforward definition.
For this reason, we use curves of precision and recall as a function of $\mathrm{T_{cls}}$  in Figure~\ref{fig:roc_alternative} to quantify the performance instead of the more used receiver operator characteristic curve, which shows the true positive rate of a classifier as a function of false positive rate. 
%This is because for an algorithm similar to GSpyNetTreeS, false positive rate is not easily defined as it depends on the entire number of negative samples. 
For a given $\mathrm{T_{cls}}$, the precision and recall are given below
\begin{equation}
	\begin{aligned}
		\mathrm{Precision} &= \frac{N_\mathrm{TP}}{N_\mathrm{TP} + N_\mathrm{FP}};\\
		\mathrm{Recall} &= \frac{N_\mathrm{TP}}{N_\mathrm{TP} + N_\mathrm{FN}},
	\end{aligned}
\end{equation}
where $N_\mathrm{TP}$, $N_\mathrm{FP}$ and $N_\mathrm{FN}$ are the number of true positives, false positives and false negatives. Precision indicates the proportion of correctly identified true positives out of all valid detections made by \texttt{GSpyNetTreeS}, reflecting its tendency to avoid false positive. Recall measures the proportion of true positives out of all actual regions of excess power, reflecting the \texttt{GSpyNetTreeS}'s ability to identify relevant regions of excess power in the images. These metrics provide an overall accuracy of the algorithm in detecting and classifying the presence of regions of excess power. 

\begin{figure}
\includegraphics[width=0.45\textwidth]{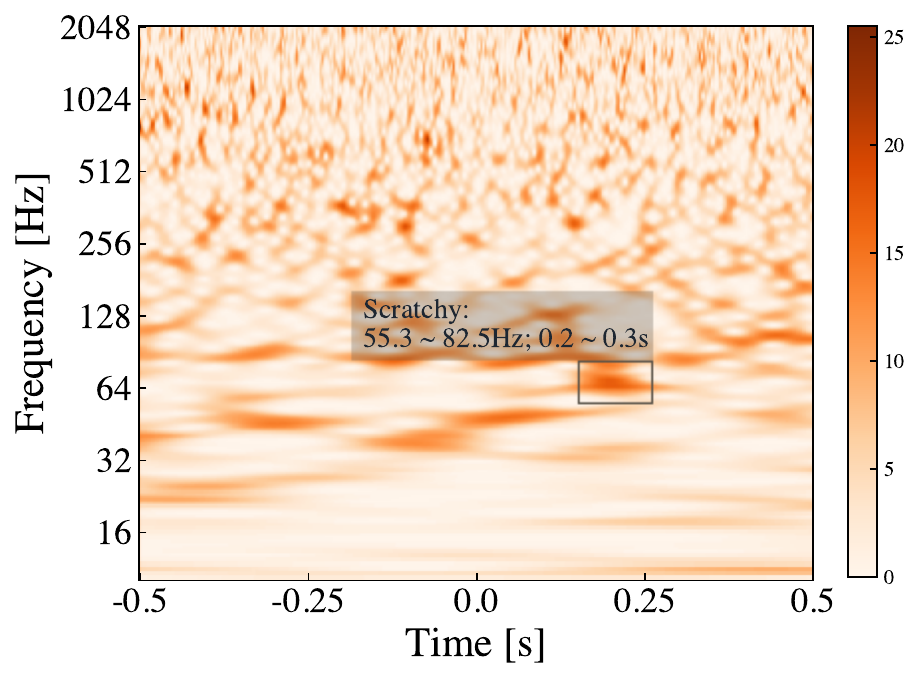}
\caption{A time-frequency representation of a one-second segment of the data collected by LIGO Livingston during \ac{O3}. A region of excess power is manually identified and classified as a Scratchy based on the definition provided in Section~\ref{sec:training}. The region of excess power is not detected by \texttt{GSpyNetTreeS} as it barely stands out from the background. 
\label{fig:weak_scratchy}}
\end{figure}

As shown in Figure~\ref{fig:roc_alternative}, both the precision and recall curves change only slightly over the range of $\mathrm{T_{cls}}$. This indicates that the performance of \texttt{GSpyNetTreeS} only depend on $\mathrm{T_{cls}}$ very weakly. This results from \texttt{GSpyNetTreeS}'s tendency to generate classification confidence $\hat{y}^{\mathrm{C}}_{ijkc}$ close to 0 or 1 and is therefore not affected by $\mathrm{T_{cls}}$ until $\mathrm{T_{cls}}$ becomes large. For reference, the $N_\mathrm{TP}$, $N_\mathrm{FP}$ and $N_\mathrm{FN}$ at $\mathrm{T_{cls}}$ equal to $0.5$ are provided in Table~\ref{table:actual_values}. 
\begin{table}[]
\centering
\begin{tabular}{llll}
\toprule
Morphology         & $N_\mathrm{TP}$ & $N_\mathrm{FP}$ & $N_\mathrm{FN}$ \\
\hline
\ac{GW}            & 1297(93.7\%)          & 16(0.1\%)              & 88(6.3\%)    \\
Blip               & 484(94.3\%)          & 10(0.2\%)            & 29(5.7\%)     \\
Low frequency blip & 428(95.3\%)          & 27(6.0\%)              & 21(4.7\%)      \\
Low frequency line & 545(82.0\%)          & 120(18.0\%)            & 122(18.0\%)     \\
Scattering         & 131(52.4\%)          & 19(7.6\%)            & 119(47.6\%)     \\
Scratchy           & 2108(87.2\%)          & 126(5.2\%)            & 311(12.8\%)    \\
\hline
\end{tabular}
\caption{The performance of \texttt{GSpyNetTreeS} quantified by the true positives $N_\mathrm{TP}$, false positives $N_\mathrm{FP}$ and false negatives $N_\mathrm{FN}$. This table assumes a $\mathrm{T_{IoU}}$ of 0.3 and a $\mathrm{T_{cls}}$ of 0.5. The percentages provided in the brackets for $N_\mathrm{TP}$ and $N_\mathrm{FN}$ are true positive rates and false negative rates. As for \texttt{GSpyNetTreeS}, there is no straightforward definition of false positive rate, the percentages provided in the brackets for $N_\mathrm{FP}$ are $N_\mathrm{FP}$ divided by the total number testing samples (see Table~\ref{table:roepn}). \label{table:actual_values}}
\end{table}
The performance for \ac{GW}, Blip and Low frequency blip is promising. For these three classes, the true positive rates are over 93\% while the false negative rates are low single digits. However, for \ac{GW}, this likely results in part from the training set exclusion of \ac{GW} signals to weak to generate excess power meeting the criteria in Section~\ref{sec:gssec}. 
%\hf{Could this be quantified? e.g. how many GW injections were `weak' (and what is the definition of weak)} 
The performance for Scratchy is poorer despite this class having the largest number of samples. This is because using our definition of region of excess power provided in Section~\ref{sec:gssec}, many glitches classified as Scratchy are small regions with energy only slightly above the background. An example is shown in Figure~\ref{fig:weak_scratchy}. Some Scratchy glitches would be considered background if a different quality factor Q was used or if the time series was slightly shifted in time, altering the background estimation. To a lesser extent, this issue also affects Low frequency lines. The performance for Scattering is impacted by both this problem and insufficient training samples. We expect a more refined definition of the region of excess power (as well as a more refined Q value) that better accounts for background variations will help improve performance.

%\begin{figure}
%    \centering
%    \subfigure[]{%
%        \includegraphics[width=\columnwidth]{Plot/Precision.pdf}
%        \label{fig:recision}
%    }\qquad
%    \subfigure[]{%   
%        \includegraphics[width=\columnwidth]{Plot/Recall.pdf}
%        \label{fig:recall}
%   }%\qquad
   %\subfloat[]{%   
   %     \includegraphics[width=\columnwidth]{make_plots/paper_plots/ft_cm.pdf}
   %     \label{fig:cmcm2}
   %}
%\caption{Curves showing the precision and recall of the model. 
%\label{fig:roc_alternative}}
%\end{figure}

Commonly used metrics for classifiers such as precision and recall are generally not correlated with how accurate the predicted bounding boxes are compared to the corresponding ground-truth bounding boxes. For example, two object detection algorithms with the same precision and recall can still perform vastly differently in terms of predicting bounding boxes accurately. 
To comprehensively assess \texttt{GSpyNetTreeS}'s performance it is necessary to understand whether the predicted bounding boxes (and associated frequency ranges and time windows of the regions of excess power) are consistent with the ground truth. For this purpose,  we use~\ac{IOU} to quantify how well the predicted bounding boxes match the ground truths.
%since we use time-frequency representation images of \ac{GW} data as inputs, 
%%the sizes and locations of predicted bounding boxes are directly related to the frequency range and time window of any glitches and \acp{GW} detected, the identification of which is a major motivation for developing \texttt{GSpyNetTreeS}. 
%It is therefore necessary to understand whether the predicted bounding boxes and therefore the frequency ranges and time windows of the regions of excess power are consistent with the ground truths. For this purpose,  we use~\ac{IOU} to quantify how well the predicted bounding boxes match the ground truths.
%To address this issue, we use \ac{IOU}. \ac{IOU} is an evaluation metric often used to quantify the performance of an object detection algorithm. \ac{IOU} is defined as the degree of overlap between a predicted bounding box and a ground truth bounding box. Its mathematical definition is given as. A value closer to 1 indicates that the predicted bounding box matches ground truth bounding box more closely. 

\begin{figure}
\gridline{
    \fig{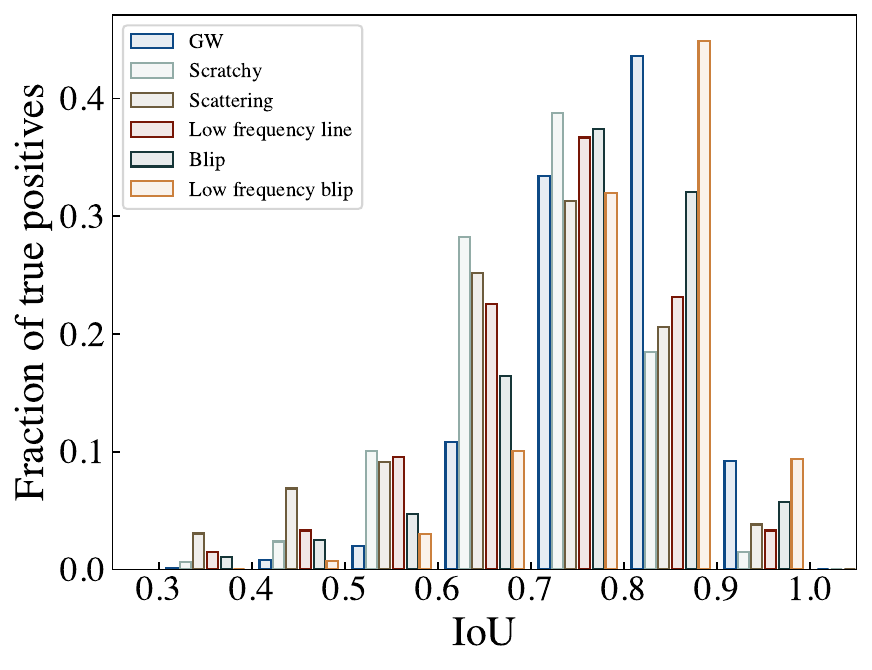}{0.45\textwidth}{(a)\label{fig:iou_dist}}
    }
\gridline{
    \fig{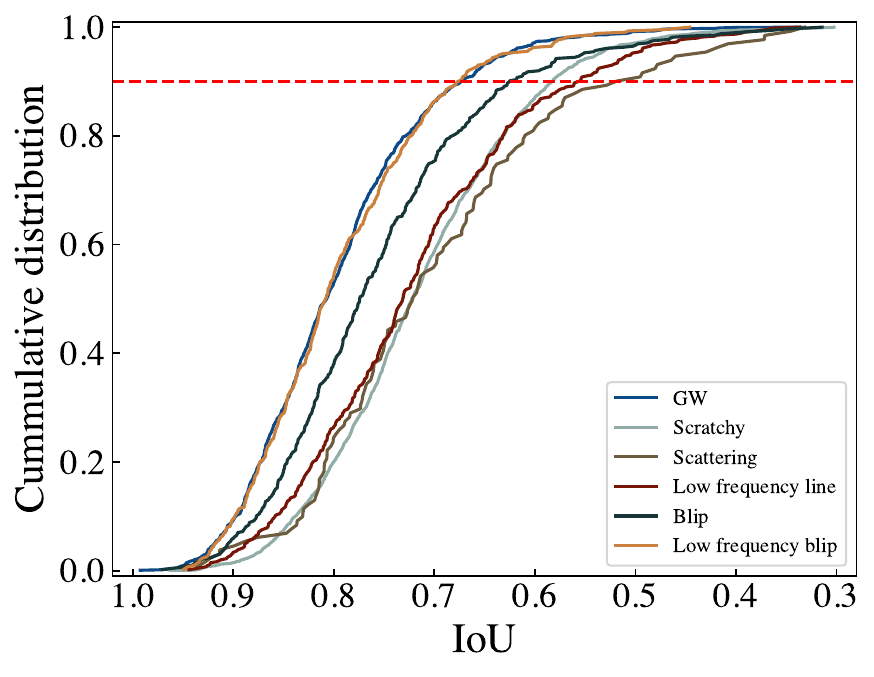}{0.45\textwidth}{(b)\label{fig:iou_commu}}
   }
\caption{The match between the predicted bounding boxes by \texttt{GSpyNetTreeS} and the ground truths for the testing samples, quantified by \ac{IOU}. The upper panel shows the distribution of \ac{IOU} for each of the classes of region of excess power. The lower panel shows the cumulative distributions. The red dashed horizontal line includes 90\% of the testing samples.
\label{fig:iou_stat}}
\end{figure}

We show in Figure~\ref{fig:iou_dist} the distribution of \ac{IOU} for the valid detections that have a corresponding ground truth box (i.e., the true positives). The distributions for all glitches and \acp{GW} are cut off at 0.3 because of the selected value of $\mathrm{T_{IoU}}$. It can be seen that for most classes, the performance is similar with the peaks of the distributions centered between 0.7 - 0.9. Figure~\ref{fig:iou_commu} shows the cumulative distributions of \ac{IOU}. The dashed horizontal line indicates that 90\% of correctly identified \ac{GW}, Scratchy, Scattering, Low frequency line, Blip, and Low frequency blip have an \ac{IOU} of larger than 0.67, 0.58, 0.53, 0.56, 0.63, and 0.68 respectively. To provide a general idea of how accurate the predicted bounding boxes represented by such values of \ac{IOU}, we calculate the \ac{IOU} for the three predicted bounding boxes in Figure~\ref{fig:groundtruth_blf_example}. For the \ac{GW}, Low frequency line and Blip,  the \acp{IOU} are 0.73, 0.85 and 0.73 respectively. Since the ground truth bounding boxes in the training samples are manually drawn, inherent variations not due to the regions of excess power or the training process are expected and will be reflected in the results. 

\section{Conclusion}\label{sec:conclusion}
In this paper, as a proof of concept, we demonstrated \texttt{GSpyNetTreeS}, a novel machine learning algorithm based on \ac{YOLO}, for the automatic identification, localization in the time and frequency domain, and classification of glitches that is robust to any \acp{GW} present.  To train \texttt{GSpyNetTreeS},  we used glitches frequently observed by the LIGO detectors including Blip, Low frequency blip, Low frequency line, Scratchy, and Scattering, and simulated \acp{GW} added to real data collected during \ac{O3}. 

We demonstrated the performance of \texttt{GSpyNetTreeS} using metrics such as precision, recall and \ac{IOU}. We showed that \texttt{GSpyNetTreeS} performs well for \ac{GW}, Blip and Low frequency blip and reasonably well for Scratchy and Low frequency line. The poor performance for Scattering suggested insufficient training samples for this type of glitches. Using the metric \ac{IOU}, we showed that for the regions of excess power that are correctly identified, \texttt{GSpyNetTreeS} can provide information such the frequency ranges and time windows that are consistent and representative of the ground truth regions of excess power. This suggests that \texttt{GSpyNetTreeS} can offer useful information for event validation and minimize human interference in the current event validation process employed by the \ac{LVK} Collaboration, making results consistent and reproducible across different \ac{GW} events. This approach is complementary to the more recently developed \texttt{GW-YOLO}, a segmentation algorithm that outputs pixel localization of \acp{GW} and glitches~\citep{Soni:2025fqu}. \texttt{GSpyNetTreeS}, which leverages the multi-label capabilities of \texttt{GSpyNetTree} to differentiate between \acp{GW} and multiple types of glitches in the same image, offers finer-grained classification of identified glitches. Additionally, \texttt{GSpyNetTreeS}'s time-frequency box output is designed to plug directly into existing glitch mitigation tools (e.g. \texttt{BayesWave}) which search an identified region for coherent signal power and incoherent glitch power. 

In the future, increasing the number of samples in training for Scattering and Low frequency line beyond what is currently available in the \texttt{GSpyNetTree} data set is expected to improve the performance of \texttt{GSpyNetTreeS} for these two types of glitches. 
Additionally, glitches with different characteristic time scales may not be fully captured by a time-frequency representation of one-second in duration. For this reason, image-based glitch classification pioneered by \texttt{Gravity Spy}~\citep{Zevin:2016qwy}, including \texttt{GSpyNetTree}~\citep{Alvarez-Lopez:2023dmv}, uses multiple time-frequency representations of different time-scales to provide the algorithms with more comprehensive information of the data~\citep{Wu_2025}. One natural step to improve \texttt{GSpyNetTreeS} therefore is to include multiple time-frequency representation images with different time-scales. Including other glitches often observed by ground-based detectors such as Tomte and Koi fish will also make the algorithm more useful and applicable. 
We expect that \texttt{GSpyNetTreeS} will be a significant advance in automating LIGO-Virgo-KAGRA's event validation process for the next observing run, allowing the collaboration to maintain its high standard of results vetting with an increased expected rate of candidate \ac{GW} detections.
%These changes are expected to enhance \texttt{GSpyNetTreeS} as an algorithm for the identification of data quality issues for LIGO-Virgo-KAGRA's future observing run, providing information both useful for real time event validations and for subsequent glitch removal and noise subtraction.  

%In the future, to improve the performance of \texttt{GSpyNetTreeS}, a more refined definition for regions of excess power that can better account for detector background variation is desired. Providing \texttt{GSpyNetTreeS} with more comprehensive information by incorporating multiple time-frequency representation images over different time-scales as well as other types of glitches observed by ground-based detectors is also expected to make \texttt{GSpyNetTreeS} more accurate for glitches of different durations. These changes are expected to allow \texttt{GSpyNetTreeS} to significantly automate LIGO-Virgo-KAGRA's event validation process for the next observing run.

\section*{Acknowledgments}
We would like to express our gratitude to Beverly Berger for their insights that improved this work. We thank members of the ML-ESTEEM collaboration, in particular Daryl Haggard, Ashish Mahabal, Audrey Durand, Renée Hložek, Flavie Lavoie-Cardinal, Nayyer Raza, Alexandre Larouche, Hadi Moazen, and Frédéric Beaupré for helpful comments and discussions. J. M. acknowledges support from the Natural
Sciences and Engineering Research Council of Canada (NSERC) Discovery Grant program, the Canada
Research Chairs (CRC) program, and IUSSTF (JC-001/2017). This material is based upon work supported by NSF’s LIGO Laboratory, which is a major facility fully funded by the National Science Foundation.

\bibliography{ObjSeg}

\begin{thebibliography}{}
\expandafter\ifx\csname natexlab\endcsname\relax\def\natexlab#1{#1}\fi
\providecommand{\url}[1]{\href{#1}{#1}}
\providecommand{\dodoi}[1]{doi:~\href{http://doi.org/#1}{\nolinkurl{#1}}}
\providecommand{\doeprint}[1]{\href{http://ascl.net/#1}{\nolinkurl{http://ascl.net/#1}}}
\providecommand{\doarXiv}[1]{\href{https://arxiv.org/abs/#1}{\nolinkurl{https://arxiv.org/abs/#1}}}

\bibitem[{Aasi {et~al.}(2015)}]{LIGOScientific:2014pky}
Aasi, J., {et~al.} 2015, Class. Quant. Grav., 32, 074001,
  \dodoi{10.1088/0264-9381/32/7/074001}

\bibitem[{Abbott {et~al.}(2020)Abbott, Abbott, Abbott, Collaboration, \& the
  Virgo~Collaboration}]{Abbott_2020}
Abbott, B.~P., Abbott, R., Abbott, Collaboration, T. L.~S., \& the
  Virgo~Collaboration. 2020, Classical and Quantum Gravity, 37, 055002,
  \dodoi{10.1088/1361-6382/ab685e}

\bibitem[{Abbott {et~al.}(2016)Abbott, Abbott, Abbott, Abernathy, Acernese,
  Ackley, Adams, Adams, Addesso, Adhikari, Adya, Affeldt, Agathos, Agatsuma,
  Aggarwal, Aguiar, Aiello, Ain, Ajith, Allen, Allocca, Altin, Anderson,
  Anderson, Arai, Arain, Araya, Arceneaux, Areeda, Arnaud, Arun, Ascenzi,
  Ashton, Ast, Aston, Astone, Aufmuth, Aulbert, Babak, Bacon, Bader, Baker,
  Baldaccini, Ballardin, Ballmer, Barayoga, Barclay, Barish, Barker, Barone,
  Barr, Barsotti, Barsuglia, Barta, Bartlett, Barton, Bartos, Bassiri, Basti,
  Batch, Baune, Bavigadda, Bazzan, Behnke, Bejger, Belczynski, Bell, Bell,
  Berger, Bergman, Bergmann, Berry, Bersanetti, Bertolini, Betzwieser, Bhagwat,
  Bhandare, Bilenko, Billingsley, Birch, Birney, Birnholtz, Biscans, Bisht,
  Bitossi, Biwer, Bizouard, Blackburn, Blair, Blair, Blair, Bloemen, Bock,
  Bodiya, Boer, Bogaert, Bogan, Bohe, Bojtos, Bond, Bondu, Bonnand, Boom, Bork,
  Boschi, Bose, Bouffanais, Bozzi, Bradaschia, Brady, Braginsky, Branchesi,
  Brau, Briant, Brillet, Brinkmann, Brisson, Brockill, Brooks, Brown, Brown,
  Brown, Buchanan, Buikema, Bulik, Bulten, Buonanno, Buskulic, Buy, Byer,
  Cabero, Cadonati, Cagnoli, Cahillane, Bustillo, Callister, Calloni, Camp,
  Cannon, Cao, Capano, Capocasa, Carbognani, Caride, Diaz, Casentini, Caudill,
  Cavagli\`a, Cavalier, Cavalieri, Cella, Cepeda, Baiardi, Cerretani, Cesarini,
  Chakraborty, Chalermsongsak, Chamberlin, Chan, Chao, Charlton,
  Chassande-Mottin, Chen, Chen, Cheng, Chincarini, Chiummo, Cho, Cho, Chow,
  Christensen, Chu, Chua, Chung, Ciani, Clara, Clark, Cleva, Coccia, Cohadon,
  Colla, Collette, Cominsky, Constancio, Conte, Conti, Cook, Corbitt, Cornish,
  Corsi, Cortese, Costa, Coughlin, Coughlin, Coulon, Countryman, Couvares,
  Cowan, Coward, Cowart, Coyne, Coyne, Craig, Creighton, Creighton, Cripe,
  Crowder, Cruise, Cumming, Cunningham, Cuoco, Canton, Danilishin, D'Antonio,
  Danzmann, Darman, Da~Silva~Costa, Dattilo, Dave, Daveloza, Davier, Davies,
  Daw, Day, De, DeBra, Debreczeni, Degallaix, De~Laurentis, Del\'eglise,
  Del~Pozzo, Denker, Dent, Dereli, Dergachev, DeRosa, De~Rosa, DeSalvo,
  Dhurandhar, D\'{\i}az, Di~Fiore, Di~Giovanni, Di~Lieto, Di~Pace, Di~Palma,
  Di~Virgilio, Dojcinoski, Dolique, Donovan, Dooley, Doravari, Douglas, Downes,
  Drago, Drever, Driggers, Du, Ducrot, Dwyer, Edo, Edwards, Effler, Eggenstein,
  Ehrens, Eichholz, Eikenberry, Engels, Essick, Etzel, Evans, Evans, Everett,
  Factourovich, Fafone, Fair, Fairhurst, Fan, Fang, Farinon, Farr, Farr,
  Favata, Fays, Fehrmann, Fejer, Feldbaum, Ferrante, Ferreira, Ferrini,
  Fidecaro, Finn, Fiori, Fiorucci, Fisher, Flaminio, Fletcher, Fong, Fournier,
  Franco, Frasca, Frasconi, Frede, Frei, Freise, Frey, Frey, Fricke, Fritschel,
  Frolov, Fulda, Fyffe, Gabbard, Gair, Gammaitoni, Gaonkar, Garufi, Gatto,
  Gaur, Gehrels, Gemme, Gendre, Genin, Gennai, George, Gergely, Germain, Ghosh,
  Ghosh, Ghosh, Giaime, Giardina, Giazotto, Gill, Glaefke, Gleason, Goetz,
  Goetz, Gondan, Gonz\'alez, Castro, Gopakumar, Gordon, Gorodetsky, Gossan,
  Gosselin, Gouaty, Graef, Graff, Granata, Grant, Gras, Gray, Greco, Green,
  Greenhalgh, Groot, Grote, Grunewald, Guidi, Guo, Gupta, Gupta, Gushwa,
  Gustafson, Gustafson, Hacker, Hall, Hall, Hammond, Haney, Hanke, Hanks,
  Hanna, Hannam, Hanson, Hardwick, Harms, Harry, Harry, Hart, Hartman, Haster,
  Haughian, Healy, Heefner, Heidmann, Heintze, Heinzel, Heitmann, Hello,
  Hemming, Hendry, Heng, Hennig, Heptonstall, Heurs, Hild, Hoak, Hodge, Hofman,
  Hollitt, Holt, Holz, Hopkins, Hosken, Hough, Houston, Howell, Hu, Huang,
  Huerta, Huet, Hughey, Husa, Huttner, Huynh-Dinh, Idrisy, Indik, Ingram, Inta,
  Isa, Isac, Isi, Islas, Isogai, Iyer, Izumi, Jacobson, Jacqmin, Jang, Jani,
  Jaranowski, Jawahar, Jim\'enez-Forteza, Johnson, Johnson-McDaniel, Jones,
  Jones, Jonker, Ju, Haris, Kalaghatgi, Kalogera, Kandhasamy, Kang, Kanner,
  Karki, Kasprzack, Katsavounidis, Katzman, Kaufer, Kaur, Kawabe, Kawazoe,
  K\'ef\'elian, Kehl, Keitel, Kelley, Kells, Kennedy, Keppel, Key,
  Khalaidovski, Khalili, Khan, Khan, Khan, Khazanov, Kijbunchoo, Kim, Kim, Kim,
  Kim, Kim, Kim, King, King, Kinzel, Kissel, Kleybolte, Klimenko, Koehlenbeck,
  Kokeyama, Koley, Kondrashov, Kontos, Koranda, Korobko, Korth, Kowalska,
  Kozak, Kringel, Krishnan, Kr\'olak, Krueger, Kuehn, Kumar, Kumar, Kuo,
  Kutynia, Kwee, Lackey, Landry, Lange, Lantz, Lasky, Lazzarini, Lazzaro,
  Leaci, Leavey, Lebigot, Lee, Lee, Lee, Lee, Lenon, Leonardi, Leong, Leroy,
  Letendre, Levin, Levine, Li, Libson, Littenberg, Lockerbie, Logue, Lombardi,
  London, Lord, Lorenzini, Loriette, Lormand, Losurdo, Lough, Lousto, Lovelace,
  L\"uck, Lundgren, Luo, Lynch, Ma, MacDonald, Machenschalk, MacInnis, Macleod,
  Maga\~na Sandoval, Magee, Mageswaran, Majorana, Maksimovic, Malvezzi, Man,
  Mandel, Mandic, Mangano, Mansell, Manske, Mantovani, Marchesoni, Marion,
  M\'arka, M\'arka, Markosyan, Maros, Martelli, Martellini, Martin, Martin,
  Martynov, Marx, Mason, Masserot, Massinger, Masso-Reid, Matichard, Matone,
  Mavalvala, Mazumder, Mazzolo, McCarthy, McClelland, McCormick, McGuire,
  McIntyre, McIver, McManus, McWilliams, Meacher, Meadors, Meidam, Melatos,
  Mendell, Mendoza-Gandara, Mercer, Merilh, Merzougui, Meshkov, Messenger,
  Messick, Meyers, Mezzani, Miao, Michel, Middleton, Mikhailov, Milano, Miller,
  Millhouse, Minenkov, Ming, Mirshekari, Mishra, Mitra, Mitrofanov,
  Mitselmakher, Mittleman, Moggi, Mohan, Mohapatra, Montani, Moore, Moore,
  Moraru, Moreno, Morriss, Mossavi, Mours, Mow-Lowry, Mueller, Mueller, Muir,
  Mukherjee, Mukherjee, Mukherjee, Mukund, Mullavey, Munch, Murphy, Murray,
  Mytidis, Nardecchia, Naticchioni, Nayak, Necula, Nedkova, Nelemans, Neri,
  Neunzert, Newton, Nguyen, Nielsen, Nissanke, Nitz, Nocera, Nolting,
  Normandin, Nuttall, Oberling, Ochsner, O'Dell, Oelker, Ogin, Oh, Oh, Ohme,
  Oliver, Oppermann, Oram, O'Reilly, O'Shaughnessy, Ott, Ottaway, Ottens,
  Overmier, Owen, Pai, Pai, Palamos, Palashov, Palomba, Pal-Singh, Pan, Pan,
  Pankow, Pannarale, Pant, Paoletti, Paoli, Papa, Paris, Parker, Pascucci,
  Pasqualetti, Passaquieti, Passuello, Patricelli, Patrick, Pearlstone,
  Pedraza, Pedurand, Pekowsky, Pele, Penn, Perreca, Pfeiffer, Phelps, Piccinni,
  Pichot, Pickenpack, Piergiovanni, Pierro, Pillant, Pinard, Pinto, Pitkin,
  Poeld, Poggiani, Popolizio, Post, Powell, Prasad, Predoi, Premachandra,
  Prestegard, Price, Prijatelj, Principe, Privitera, Prix, Prodi, Prokhorov,
  Puncken, Punturo, Puppo, P\"urrer, Qi, Qin, Quetschke, Quintero,
  Quitzow-James, Raab, Rabeling, Radkins, Raffai, Raja, Rakhmanov, Ramet,
  Rapagnani, Raymond, Razzano, Re, Read, Reed, Regimbau, Rei, Reid, Reitze,
  Rew, Reyes, Ricci, Riles, Robertson, Robie, Robinet, Rocchi, Rolland,
  Rollins, Roma, Romano, Romano, Romanov, Romie, Rosi\ifmmode~\acute{n}\else
  \'{n}\fi{}ska, Rowan, R\"udiger, Ruggi, Ryan, Sachdev, Sadecki, Sadeghian,
  Salconi, Saleem, Salemi, Samajdar, Sammut, Sampson, Sanchez, Sandberg,
  Sandeen, Sanders, Sanders, Sassolas, Sathyaprakash, Saulson, Sauter, Savage,
  Sawadsky, Schale, Schilling, Schmidt, Schmidt, Schnabel, Schofield,
  Sch\"onbeck, Schreiber, Schuette, Schutz, Scott, Scott, Sellers, Sengupta,
  Sentenac, Sequino, Sergeev, Serna, Setyawati, Sevigny, Shaddock, Shaffer,
  Shah, Shahriar, Shaltev, Shao, Shapiro, Shawhan, Sheperd, Shoemaker,
  Shoemaker, Siellez, Siemens, Sigg, Silva, Simakov, Singer, Singer, Singh,
  Singh, Singhal, Sintes, Slagmolen, Smith, Smith, Smith, Smith, Son, Sorazu,
  Sorrentino, Souradeep, Srivastava, Staley, Steinke, Steinlechner,
  Steinlechner, Steinmeyer, Stephens, Stevenson, Stone, Strain, Straniero,
  Stratta, Strauss, Strigin, Sturani, Stuver, Summerscales, Sun, Sutton,
  Swinkels, Szczepa\ifmmode~\acute{n}\else \'{n}\fi{}czyk, Tacca, Talukder,
  Tanner, T\'apai, Tarabrin, Taracchini, Taylor, Theeg, Thirugnanasambandam,
  Thomas, Thomas, Thomas, Thorne, Thorne, Thrane, Tiwari, Tiwari, Tokmakov,
  Tomlinson, Tonelli, Torres, Torrie, T\"oyr\"a, Travasso, Traylor, Trifir\`o,
  Tringali, Trozzo, Tse, Turconi, Tuyenbayev, Ugolini, Unnikrishnan, Urban,
  Usman, Vahlbruch, Vajente, Valdes, Vallisneri, van Bakel, van Beuzekom,
  van~den Brand, Van Den~Broeck, Vander-Hyde, van~der Schaaf, van Heijningen,
  van Veggel, Vardaro, Vass, Vas\'uth, Vaulin, Vecchio, Vedovato, Veitch,
  Veitch, Venkateswara, Verkindt, Vetrano, Vicer\'e, Vinciguerra, Vine, Vinet,
  Vitale, Vo, Vocca, Vorvick, Voss, Vousden, Vyatchanin, Wade, Wade, Wade,
  Waldman, Walker, Wallace, Walsh, Wang, Wang, Wang, Wang, Wang, Ward, Ward,
  Warner, Was, Weaver, Wei, Weinert, Weinstein, Weiss, Welborn, Wen,
  We\ss{}els, Westphal, Wette, Whelan, Whitcomb, White, Whiting, Wiesner,
  Wilkinson, Willems, Williams, Williams, Williamson, Willis, Willke, Wimmer,
  Winkelmann, Winkler, Wipf, Wiseman, Wittel, Woan, Worden, Wright, Wu, Yablon,
  Yakushin, Yam, Yamamoto, Yancey, Yap, Yu, Yvert, Zadro\ifmmode~\dot{z}\else
  \.{z}\fi{}ny, Zangrando, Zanolin, Zendri, Zevin, Zhang, Zhang, Zhang, Zhang,
  Zhao, Zhou, Zhou, Zhu, Zucker, Zuraw, \& Zweizig}]{PhysRevLett.116.061102}
Abbott, B.~P., Abbott, R., Abbott, T.~D., {et~al.} 2016, Phys. Rev. Lett., 116,
  061102, \dodoi{10.1103/PhysRevLett.116.061102}

\bibitem[{Abbott {et~al.}(2019)}]{LIGOScientific:2018mvr}
Abbott, B.~P., {et~al.} 2019, Phys. Rev. X, 9, 031040,
  \dodoi{10.1103/PhysRevX.9.031040}

\bibitem[{Abbott {et~al.}(2021{\natexlab{a}})}]{LIGOScientific:2020ibl}
Abbott, R., {et~al.} 2021{\natexlab{a}}, Phys. Rev. X, 11, 021053,
  \dodoi{10.1103/PhysRevX.11.021053}

\bibitem[{Abbott {et~al.}(2021{\natexlab{b}})}]{KAGRA:2021tnv}
---. 2021{\natexlab{b}}, Phys. Rev. D, 104, 122004,
  \dodoi{10.1103/PhysRevD.104.122004}

\bibitem[{Abbott {et~al.}(2022{\natexlab{a}})}]{KAGRA:2022dwb}
---. 2022{\natexlab{a}}, Phys. Rev. D, 106, 102008,
  \dodoi{10.1103/PhysRevD.106.102008}

\bibitem[{Abbott {et~al.}(2023)}]{KAGRA:2021vkt}
---. 2023, Phys. Rev. X, 13, 041039, \dodoi{10.1103/PhysRevX.13.041039}

\bibitem[{Abbott {et~al.}(2022{\natexlab{b}})Abbott, Buffaz, Vieira, Cabero,
  Haggard, Mahabal, \& McIver}]{Abbott_2022}
Abbott, T.~C., Buffaz, E., Vieira, N., {et~al.} 2022{\natexlab{b}}, The
  Astrophysical Journal, 927, 232, \dodoi{10.3847/1538-4357/ac5019}

\bibitem[{Acernese {et~al.}(2015)}]{VIRGO:2014yos}
Acernese, F., {et~al.} 2015, Class. Quant. Grav., 32, 024001,
  \dodoi{10.1088/0264-9381/32/2/024001}

\bibitem[{Alif \& Hussain(2024)}]{alif2024yolov1}
Alif, M. A.~R., \& Hussain, M. 2024, arXiv preprint arXiv:2406.10139

\bibitem[{{All\'en\'e} {et~al.}(2025){All\'en\'e}, Aubin, Bentara, Buskulic,
  Guidi, Juste, Lethuillier, Marion, Mobilia, Mours, Ouzriat, Sainrat, \&
  Sordini}]{Allene_2025}
{All\'en\'e}, C., Aubin, F., Bentara, I., {et~al.} 2025, Classical and Quantum
  Gravity, 42, 105009, \dodoi{10.1088/1361-6382/add234}

\bibitem[{Alvarez-Lopez {et~al.}(2024)Alvarez-Lopez, Liyanage, Ding, Ng, \&
  McIver}]{Alvarez-Lopez:2023dmv}
Alvarez-Lopez, S., Liyanage, A., Ding, J., Ng, R., \& McIver, J. 2024, Class.
  Quant. Grav., 41, 085007, \dodoi{10.1088/1361-6382/ad2194}

\bibitem[{Areeda {et~al.}(2017)Areeda, Smith, Lundgren, Maros, Macleod, \&
  Zweizig}]{Areeda:2016mee}
Areeda, J.~S., Smith, J.~R., Lundgren, A.~P., {et~al.} 2017, Astron. Comput.,
  18, 27, \dodoi{10.1016/j.ascom.2017.01.003}

\bibitem[{Aso {et~al.}(2013)Aso, Michimura, Somiya, Ando, Miyakawa, Sekiguchi,
  Tatsumi, \& Yamamoto}]{Aso:2013eba}
Aso, Y., Michimura, Y., Somiya, K., {et~al.} 2013, Phys. Rev. D, 88, 043007,
  \dodoi{10.1103/PhysRevD.88.043007}

\bibitem[{Berger(2018)}]{Berger_2018}
Berger, B.~K. 2018, Journal of Physics: Conference Series, 957, 012004,
  \dodoi{10.1088/1742-6596/957/1/012004}

\bibitem[{Biswas {et~al.}(2013)Biswas, Blackburn, Cao, Essick, Hodge,
  Katsavounidis, Kim, Kim, Le~Bigot, Lee, Oh, Oh, Son, Tao, Vaulin, \&
  Wang}]{PhysRevD.88.062003}
Biswas, R., Blackburn, L., Cao, J., {et~al.} 2013, Phys. Rev. D, 88, 062003,
  \dodoi{10.1103/PhysRevD.88.062003}

\bibitem[{Cabero {et~al.}(2020)Cabero, Mahabal, \& McIver}]{Cabero:2020eik}
Cabero, M., Mahabal, A., \& McIver, J. 2020, Astrophys. J. Lett., 904, L9,
  \dodoi{10.3847/2041-8213/abc5b5}

\bibitem[{Canton {et~al.}(2014)Canton, Bhagwat, Dhurandhar, \&
  Lundgren}]{Canton:2013joa}
Canton, T.~D., Bhagwat, S., Dhurandhar, S.~V., \& Lundgren, A. 2014, Class.
  Quant. Grav., 31, 015016, \dodoi{10.1088/0264-9381/31/1/015016}

\bibitem[{Chan {et~al.}(2020)Chan, Heng, \& Messenger}]{Chan:2019fuz}
Chan, M.~L., Heng, I.~S., \& Messenger, C. 2020, Phys. Rev. D, 102, 043022,
  \dodoi{10.1103/PhysRevD.102.043022}

\bibitem[{Chan {et~al.}(2024)Chan, McIver, Mahabal, Messick, Haggard, Raza,
  Lecoeuche, Sutton, Ewing, Di~Renzo, Cabero, Ng, Coughlin, Ghosh, \&
  Godwin}]{Chan_2024}
Chan, M.~L., McIver, J., Mahabal, A., {et~al.} 2024, The Astrophysical Journal,
  972, 50, \dodoi{10.3847/1538-4357/ad496a}

\bibitem[{Chatterjee {et~al.}(2023)Chatterjee, Kovalam, Wen, Beveridge,
  Diakogiannis, \& Vinsen}]{Chatterjee:2022ggk}
Chatterjee, C., Kovalam, M., Wen, L., {et~al.} 2023, Astrophys. J., 959, 42,
  \dodoi{10.3847/1538-4357/ad08b7}

\bibitem[{Chatterji {et~al.}(2004)Chatterji, Blackburn, Martin, \&
  Katsavounidis}]{Chatterji:2004qg}
Chatterji, S., Blackburn, L., Martin, G., \& Katsavounidis, E. 2004, Class.
  Quant. Grav., 21, S1809, \dodoi{10.1088/0264-9381/21/20/024}

\bibitem[{Cheng {et~al.}(2024)Cheng, Song, Ge, Liu, Wang, \&
  Shan}]{cheng2024yoloworldrealtimeopenvocabularyobject}
Cheng, T., Song, L., Ge, Y., {et~al.} 2024, YOLO-World: Real-Time
  Open-Vocabulary Object Detection.
\newblock \doarXiv{2401.17270}

\bibitem[{Chu {et~al.}(2022)Chu, Kovalam, Wen, Slaven-Blair, Bosveld, Chen,
  Clearwater, Codoreanu, Du, Guo, Guo, Kim, Li, Oloworaran, Panther, Powell,
  Sengupta, Wette, \& Zhu}]{PhysRevD.105.024023}
Chu, Q., Kovalam, M., Wen, L., {et~al.} 2022, Phys. Rev. D, 105, 024023,
  \dodoi{10.1103/PhysRevD.105.024023}

\bibitem[{Cornish \& Littenberg(2015)}]{Cornish_2015}
Cornish, N.~J., \& Littenberg, T.~B. 2015, Classical and Quantum Gravity, 32,
  135012, \dodoi{10.1088/0264-9381/32/13/135012}

\bibitem[{Cuoco {et~al.}(2021)}]{Cuoco:2020ogp}
Cuoco, E., {et~al.} 2021, Mach. Learn. Sci. Tech., 2, 011002,
  \dodoi{10.1088/2632-2153/abb93a}

\bibitem[{Dal~Canton {et~al.}(2021)Dal~Canton, Nitz, Gadre, Cabourn~Davies,
  Villa-Ortega, Dent, Harry, \& Xiao}]{Dal_Canton_2021}
Dal~Canton, T., Nitz, A.~H., Gadre, B., {et~al.} 2021, The Astrophysical
  Journal, 923, 254, \dodoi{10.3847/1538-4357/ac2f9a}

\bibitem[{Davis {et~al.}(2022)Davis, Littenberg, Romero-Shaw, Millhouse,
  McIver, Di~Renzo, \& Ashton}]{Davis:2022ird}
Davis, D., Littenberg, T.~B., Romero-Shaw, I.~M., {et~al.} 2022, Class. Quant.
  Grav., 39, 245013, \dodoi{10.1088/1361-6382/aca238}

\bibitem[{Davis {et~al.}(2021{\natexlab{a}})Davis, Areeda, Berger, Bruntz,
  Effler, Essick, Fisher, Godwin, Goetz, Helmling-Cornell, Hughey,
  Katsavounidis, Lundgren, Macleod, Márka, Massinger, Matas, McIver, Mo,
  Mogushi, Nguyen, Nuttall, Schofield, Shoemaker, Soni, Stuver, Urban, Valdes,
  Walker, Abbott, Adams, Adhikari, Ananyeva, Appert, Arai, Asali, Aston,
  Austin, Baer, Ball, Ballmer, Banagiri, Barker, Barschaw, Barsotti, Bartlett,
  Betzwieser, Beda, Bhattacharjee, Bidler, Billingsley, Biscans, Blair, Blair,
  Bode, Booker, Bork, Bramley, Brooks, Brown, Buikema, Cahillane, Callister,
  Caneva~Santoro, Cannon, Carlin, Chandra, Chen, Christensen, Ciobanu, Clara,
  Compton, Cooper, Corley, Coughlin, Countryman, Covas, Coyne, Crowder,
  Dal~Canton, Danila, Datrier, Davies, Dent, Didio, Di~Fronzo, Dooley,
  Driggers, Dupej, Dwyer, Etzel, Evans, Evans, Fairhurst, Feicht,
  Fernandez-Galiana, Frey, Fritschel, Frolov, Fulda, Fyffe, Gadre, Giaime,
  Giardina, González, Gras, Gray, Gray, Green, Gupta, Gustafson, Gustafson,
  Hanks, Hanson, Hardwick, Harry, Hasskew, Heintze, Heinzel, Holland, Hollows,
  Hoy, Hughey, Jadhav, Janssens, Johns, Jones, Kandhasamy, Karki, Kasprzack,
  Kawabe, Keitel, Kijbunchoo, Kim, King, Kissel, Kulkarni, Kumar, Landry, Lane,
  Lantz, Laxen, Lecoeuche, Leviton, Liu, Lormand, Macas, Macedo, MacInnis,
  Mandic, Mansell, Márka, Martinez, Martinovic, Martynov, Mason, Matichard,
  Mavalvala, McCarthy, McClelland, McCormick, McCuller, McIsaac, McRae,
  Mendell, Merfeld, Merilh, Meyers, Meylahn, Michaloliakos, Middleton, Mills,
  Mistry, Mittleman, Moreno, Mow-Lowry, Mozzon, Mueller, Mukund, Mullavey,
  Muth, Nelson, Neunzert, Nichols, Nitoglia, Oberling, Oh, Oh, Oram, Ormiston,
  Ormsby, Osthelder, Ottaway, Overmier, Pai, Palamos, Pannarale, Parker,
  Patane, Patel, Payne, Pele, Penhorwood, Perez, Phukon, Pillas, Pirello,
  Radkins, Ramirez, Richardson, Riles, Rink, Robertson, Rollins, Romel, Romie,
  Ross, Ryan, Sadecki, Sakellariadou, Sanchez, Sanchez, Sandles, Saravanan,
  Savage, Schaetzl, Schnabel, Schwartz, Sellers, Shaffer, Sigg, Sintes,
  Slagmolen, Smith, Soni, Sorazu, Spencer, Strain, Strom, Sun, Szczepańczyk,
  Tasson, Tenorio, Thomas, Thomas, Thorne, Toland, Torrie, Tran, Traylor,
  Trevor, Tse, Vajente, van Remortel, Vander-Hyde, Vargas, Veitch, Veitch,
  Venkateswara, Venugopalan, Viets, Villa-Ortega, Vo, Vorvick, Wade, Wallace,
  Ward, Warner, Weaver, Weinstein, Weiss, Wette, White, White, Whittle,
  Williamson, Willke, Wipf, Xiao, Xu, Yamamoto, Yu, Yu, Zhang, Zheng, Zucker,
  \& Zweizig}]{Davis_2021}
Davis, D., Areeda, J.~S., Berger, B.~K., {et~al.} 2021{\natexlab{a}}, Classical
  and Quantum Gravity, 38, 135014, \dodoi{10.1088/1361-6382/abfd85}

\bibitem[{Davis {et~al.}(2021{\natexlab{b}})}]{LIGO:2021ppb}
Davis, D., {et~al.} 2021{\natexlab{b}}, Class. Quant. Grav., 38, 135014,
  \dodoi{10.1088/1361-6382/abfd85}

\bibitem[{Essick {et~al.}(2013)Essick, Blackburn, \&
  Katsavounidis}]{Essick_2013}
Essick, R., Blackburn, L., \& Katsavounidis, E. 2013, Classical and Quantum
  Gravity, 30, 155010, \dodoi{10.1088/0264-9381/30/15/155010}

\bibitem[{Ewing {et~al.}(2024)Ewing, Huxford, Singh, Tsukada, Hanna, Huang,
  Joshi, Li, Magee, Messick, Pace, Ray, Sachdev, Sakon, Tapia, Adhicary, Baral,
  Baylor, Cannon, Caudill, Chaudhary, Coughlin, Cousins, Creighton, Essick,
  Fong, George, Godwin, Harada, Kennington, Kuwahara, Meacher, Morisaki,
  Mukherjee, Niu, Posnansky, Toivonen, Tsutsui, Ueno, Viets, Wade, Wade, \&
  Waratkar}]{PhysRevD.109.042008}
Ewing, B., Huxford, R., Singh, D., {et~al.} 2024, Phys. Rev. D, 109, 042008,
  \dodoi{10.1103/PhysRevD.109.042008}

\bibitem[{Gabbard {et~al.}(2022)Gabbard, Messenger, Heng, Tonolini, \&
  Murray-Smith}]{Gabbard:2019rde}
Gabbard, H., Messenger, C., Heng, I.~S., Tonolini, F., \& Murray-Smith, R.
  2022, Nature Phys., 18, 112, \dodoi{10.1038/s41567-021-01425-7}

\bibitem[{Gabbard {et~al.}(2018)Gabbard, Williams, Hayes, \&
  Messenger}]{PhysRevLett.120.141103}
Gabbard, H., Williams, M., Hayes, F., \& Messenger, C. 2018, Phys. Rev. Lett.,
  120, 141103, \dodoi{10.1103/PhysRevLett.120.141103}

\bibitem[{Ghonge {et~al.}(2024)Ghonge, Brandt, Sullivan, Millhouse,
  Chatziioannou, Clark, Littenberg, Cornish, Hourihane, \&
  Cadonati}]{PhysRevD.110.122002}
Ghonge, S., Brandt, J., Sullivan, J.~M., {et~al.} 2024, Phys. Rev. D, 110,
  122002, \dodoi{10.1103/PhysRevD.110.122002}

\bibitem[{Glanzer {et~al.}(2021)Glanzer, Banagari, Coughlin, Zevin, Bahaadini,
  Rohani, Allen, Berry, Crowston, Harandi, Jackson, Kalogera, Katsaggelos,
  Noroozi, Osterlund, Patane, Smith, Soni, \& Trouille}]{glanzer_2021_5649212}
Glanzer, J., Banagari, S., Coughlin, S., {et~al.} 2021, Gravity Spy Machine
  Learning Classifications of LIGO Glitches from Observing Runs O1, O2, O3a,
  and O3b, v1.0.0,  Zenodo, \dodoi{10.5281/zenodo.5649212}

\bibitem[{Green {et~al.}(2020)Green, Simpson, \& Gair}]{Green:2020hst}
Green, S.~R., Simpson, C., \& Gair, J. 2020, Phys. Rev. D, 102, 104057,
  \dodoi{10.1103/PhysRevD.102.104057}

\bibitem[{Heaton(2018)}]{heaton2018ian}
Heaton, J. 2018, Genetic Programming and Evolvable Machines, 19, 305

\bibitem[{Hourihane {et~al.}(2022)Hourihane, Chatziioannou, Wijngaarden, Davis,
  Littenberg, \& Cornish}]{PhysRevD.106.042006}
Hourihane, S., Chatziioannou, K., Wijngaarden, M., {et~al.} 2022, Phys. Rev. D,
  106, 042006, \dodoi{10.1103/PhysRevD.106.042006}

\bibitem[{Husa {et~al.}(2016)Husa, Khan, Hannam, P\"urrer, Ohme, Forteza, \&
  Boh\'e}]{PhysRevD.93.044006}
Husa, S., Khan, S., Hannam, M., {et~al.} 2016, Phys. Rev. D, 93, 044006,
  \dodoi{10.1103/PhysRevD.93.044006}

\bibitem[{Jarov {et~al.}(2023)Jarov, Thiele, Soni, Ding, McIver, Ng, Hatoya, \&
  Davis}]{Jarov:2023qpt}
Jarov, S., Thiele, S., Soni, S., {et~al.} 2023.
\newblock \doarXiv{2307.15867}

\bibitem[{Jiang {et~al.}(2022)Jiang, Ergu, Liu, Cai, \& Ma}]{JIANG20221066}
Jiang, P., Ergu, D., Liu, F., Cai, Y., \& Ma, B. 2022, Procedia Computer
  Science, 199, 1066, \dodoi{https://doi.org/10.1016/j.procs.2022.01.135}

\bibitem[{Khan {et~al.}(2016)Khan, Husa, Hannam, Ohme, P\"urrer, Forteza, \&
  Boh\'e}]{PhysRevD.93.044007}
Khan, S., Husa, S., Hannam, M., {et~al.} 2016, Phys. Rev. D, 93, 044007,
  \dodoi{10.1103/PhysRevD.93.044007}

\bibitem[{Khanam \& Hussain(2024)}]{khanam2024yolov11overviewkeyarchitectural}
Khanam, R., \& Hussain, M. 2024, YOLOv11: An Overview of the Key Architectural
  Enhancements.
\newblock \doarXiv{2410.17725}

\bibitem[{Koyama {et~al.}(2024)Koyama, Sakai, Sasaoka, Dominguez, Somiya, Omae,
  Terada, Meyer-Conde, \& Takahashi}]{Koyama_2024}
Koyama, N., Sakai, Y., Sasaoka, S., {et~al.} 2024, Machine Learning: Science
  and Technology, 5, 035028, \dodoi{10.1088/2632-2153/ad6391}

\bibitem[{Langendorff {et~al.}(2023)Langendorff, Kolmus, Janquart, \& Van
  Den~Broeck}]{Langendorff:2022fzq}
Langendorff, J., Kolmus, A., Janquart, J., \& Van Den~Broeck, C. 2023, Phys.
  Rev. Lett., 130, 171402, \dodoi{10.1103/PhysRevLett.130.171402}

\bibitem[{{LIGO Scientific Collaboration}(2018)}]{lalsuite}
{LIGO Scientific Collaboration}. 2018, {LIGO} {A}lgorithm {L}ibrary -
  {LALS}uite, free software (GPL), \dodoi{10.7935/GT1W-FZ16}

\bibitem[{Macas {et~al.}(2022)Macas, Pooley, Nuttall, Davis, Dyer, Lecoeuche,
  Lyman, McIver, \& Rink}]{PhysRevD.105.103021}
Macas, R., Pooley, J., Nuttall, L.~K., {et~al.} 2022, Phys. Rev. D, 105,
  103021, \dodoi{10.1103/PhysRevD.105.103021}

\bibitem[{Mishra {et~al.}(2022)}]{Mishra:2022ott}
Mishra, T., {et~al.} 2022, Phys. Rev. D, 105, 083018,
  \dodoi{10.1103/PhysRevD.105.083018}

\bibitem[{Murali \& Lumley(2023)}]{PhysRevD.108.043024}
Murali, C., \& Lumley, D. 2023, Phys. Rev. D, 108, 043024,
  \dodoi{10.1103/PhysRevD.108.043024}

\bibitem[{Pankow {et~al.}(2018)Pankow, Chatziioannou, Chase, Littenberg, Evans,
  McIver, Cornish, Haster, Kanner, Raymond, Vitale, \&
  Zimmerman}]{PhysRevD.98.084016}
Pankow, C., Chatziioannou, K., Chase, E.~A., {et~al.} 2018, Phys. Rev. D, 98,
  084016, \dodoi{10.1103/PhysRevD.98.084016}

\bibitem[{Payne {et~al.}(2022)Payne, Hourihane, Golomb, Udall, Davis, \&
  Chatziioannou}]{PhysRevD.106.104017}
Payne, E., Hourihane, S., Golomb, J., {et~al.} 2022, Phys. Rev. D, 106, 104017,
  \dodoi{10.1103/PhysRevD.106.104017}

\bibitem[{Powell(2018)}]{Powell:2018csz}
Powell, J. 2018, Class. Quant. Grav., 35, 155017,
  \dodoi{10.1088/1361-6382/aacf18}

\bibitem[{Powell {et~al.}(2023)Powell, Sun, Gereb, Lasky, \&
  Dollmann}]{Powell:2022pcg}
Powell, J., Sun, L., Gereb, K., Lasky, P.~D., \& Dollmann, M. 2023, Class.
  Quant. Grav., 40, 035006, \dodoi{10.1088/1361-6382/acb038}

\bibitem[{Raza {et~al.}(2024)Raza, Chan, Haggard, Mahabal, McIver, Abbott,
  Buffaz, \& Vieira}]{Raza:2023gyv}
Raza, N., Chan, M.~L., Haggard, D., {et~al.} 2024, Astrophys. J., 963, 98,
  \dodoi{10.3847/1538-4357/ad13ea}

\bibitem[{Redmon {et~al.}(2016)Redmon, Divvala, Girshick, \&
  Farhadi}]{redmon2016lookonceunifiedrealtime}
Redmon, J., Divvala, S., Girshick, R., \& Farhadi, A. 2016, You Only Look Once:
  Unified, Real-Time Object Detection.
\newblock \doarXiv{1506.02640}

\bibitem[{Redmon \& Farhadi(2018)}]{redmon2018yolov3incrementalimprovement}
Redmon, J., \& Farhadi, A. 2018, YOLOv3: An Incremental Improvement.
\newblock \doarXiv{1804.02767}

\bibitem[{Rezatofighi {et~al.}(2019)Rezatofighi, Tsoi, Gwak, Sadeghian, Reid,
  \& Savarese}]{rezatofighi2019generalizedintersectionunionmetric}
Rezatofighi, H., Tsoi, N., Gwak, J., {et~al.} 2019, Generalized Intersection
  over Union: A Metric and A Loss for Bounding Box Regression.
\newblock \doarXiv{1902.09630}

\bibitem[{Skliris {et~al.}(2024)Skliris, Norman, \& Sutton}]{Skliris:2020qax}
Skliris, V., Norman, M. R.~K., \& Sutton, P.~J. 2024, Phys. Rev. D, 110,
  104034, \dodoi{10.1103/PhysRevD.110.104034}

\bibitem[{Soni {et~al.}(2025{\natexlab{a}})Soni, Mukund, \&
  Katsavounidis}]{Soni:2025fqu}
Soni, S., Mukund, N., \& Katsavounidis, E. 2025{\natexlab{a}}.
\newblock \doarXiv{2508.17399}

\bibitem[{Soni {et~al.}(2025{\natexlab{b}})Soni, Berger, Davis, Di~Renzo,
  Effler, Ferreira, Glanzer, Goetz, González, Helmling-Cornell, Hughey,
  Huxford, Mannix, Mo, Nandi, Neunzert, Nichols, Pham, Renzini, Schofield,
  Stuver, Trevor, Álvarez López, Beda, Berry, Bhuiyan, Blagg, Bruntz, Callos,
  Chan, Charlton, Christensen, Connolly, Dhatri, Ding, Garg, Holley-Bockelmann,
  Hourihane, Jani, Janssens, Jarov, Knee, Lattal, Lecoeuche, Littenberg,
  Liyanage, Lott, Macas, Malakar, McGowan, McIver, Millhouse, Nuttall, Nykamp,
  Ota, Rawcliffe, Scully, Tasson, Tejera, Thiele, Udall, Winborn, Yarbrough,
  Zhang, Zheng, Abbott, Abouelfettouh, Adhikari, Ananyeva, Appert, Arai,
  Aritomi, Aston, Ball, Ballmer, Barker, Barsotti, Betzwieser, Billingsley,
  Biscans, Bode, Bonilla, Bossilkov, Branch, Brooks, Brown, Bryant, Cahillane,
  Cao, Capote, Clara, Collins, Compton, Cottingham, Coyne, Crouch, Csizmazia,
  Cullen, Dartez, Demos, Dohmen, Driggers, Dwyer, Ejlli, Etzel, Evans, Feicht,
  Frey, Frischhertz, Fritschel, Frolov, Fulda, Fyffe, Ganapathy, Gateley,
  Giaime, Giardina, Goetz, Goodwin-Jones, Gras, Gray, Griffith, Grote, Guidry,
  Hall, Hanks, Hanson, Heintze, Holland, Hoyland, Huang, Inoue, James,
  Jennings, Jia, Karat, Karki, Kasprzack, Kawabe, Kijbunchoo, King, Kissel,
  Komori, Kontos, Kumar, Kuns, Landry, Lantz, Laxen, Lee, Lesovsky, Llamas,
  Lormand, Loughlin, MacInnis, Makarem, Mansell, Martin, Mason, Matichard,
  Mavalvala, Maxwell, McCarrol, McCarthy, McClelland, McCormick, McCuller,
  McRae, Mera, Merilh, Meylahn, Mittleman, Moraru, Moreno, Mullavey, Nakano,
  Nelson, Notte, Oberling, O’Hanlon, Osthelder, Ottaway, Overmier, Parker,
  Pele, Pham, Pirello, Quetschke, Ramirez, Reyes, Richardson, Robinson,
  Rollins, Romel, Romie, Ross, Ryan, Sadecki, Sanchez, Sanchez, Sanchez,
  Savage, Schaetzl, Schiworski, Schnabel, Schwartz, Sellers, Shaffer, Short,
  Sigg, Slagmolen, Soike, Srivastava, Sun, Tanner, Thomas, Thomas, Thorne,
  Torrie, Traylor, Ubhi, Vajente, Vanosky, Vecchio, Veitch, Vibhute, von Reis,
  Warner, Weaver, Weiss, Whittle, Willke, Wipf, Xu, Yamamoto, Zhang, \&
  Zucker}]{LIGO:2024kkz}
Soni, S., Berger, B.~K., Davis, D., {et~al.} 2025{\natexlab{b}}, Classical and
  Quantum Gravity, 42, 085016, \dodoi{10.1088/1361-6382/adc4b6}

\bibitem[{Terven {et~al.}(2023)Terven, Córdova-Esparza, \&
  Romero-González}]{Terven_2023}
Terven, J., Córdova-Esparza, D.-M., \& Romero-González, J.-A. 2023, Machine
  Learning and Knowledge Extraction, 5, 1680–1716,
  \dodoi{10.3390/make5040083}

\bibitem[{Tsukada {et~al.}(2023)Tsukada, Joshi, Adhicary, George, Guimaraes,
  Hanna, Magee, Zimmerman, Baral, Baylor, Cannon, Caudill, Cousins, Creighton,
  Ewing, Fong, Godwin, Harada, Huang, Huxford, Kennington, Kuwahara, Li,
  Meacher, Messick, Morisaki, Mukherjee, Niu, Pace, Posnansky, Ray, Sachdev,
  Sakon, Singh, Tapia, Tsutsui, Ueno, Viets, Wade, \&
  Wade}]{PhysRevD.108.043004}
Tsukada, L., Joshi, P., Adhicary, S., {et~al.} 2023, Phys. Rev. D, 108, 043004,
  \dodoi{10.1103/PhysRevD.108.043004}

\bibitem[{Udall {et~al.}(2025)Udall, Hourihane, Miller, Davis, Chatziioannou,
  Isi, \& Deshong}]{Udall:2024ovp}
Udall, R., Hourihane, S., Miller, S., {et~al.} 2025, Phys. Rev. D, 111, 024046,
  \dodoi{10.1103/PhysRevD.111.024046}

\bibitem[{Usman {et~al.}(2016)Usman, Nitz, Harry, Biwer, Brown, Cabero, Capano,
  Canton, Dent, Fairhurst, Kehl, Keppel, Krishnan, Lenon, Lundgren, Nielsen,
  Pekowsky, Pfeiffer, Saulson, West, \& Willis}]{Usman_2016}
Usman, S.~A., Nitz, A.~H., Harry, I.~W., {et~al.} 2016, Classical and Quantum
  Gravity, 33, 215004, \dodoi{10.1088/0264-9381/33/21/215004}

\bibitem[{Vazsonyi \& Davis(2023)}]{Vazsonyi:2022jul}
Vazsonyi, L., \& Davis, D. 2023, Class. Quant. Grav., 40, 035008,
  \dodoi{10.1088/1361-6382/acafd2}

\bibitem[{Wang {et~al.}(2024)Wang, Liao, {et~al.}}]{wang2024yolov1}
Wang, C.-Y., Liao, H.-Y.~M., {et~al.} 2024, APSIPA Transactions on Signal and
  Information Processing, 13

\bibitem[{Wu {et~al.}(2025)Wu, Zevin, Berry, Crowston, Østerlund, Doctor,
  Banagiri, Jackson, Kalogera, \& Katsaggelos}]{Wu_2025}
Wu, Y., Zevin, M., Berry, C. P.~L., {et~al.} 2025, Classical and Quantum
  Gravity, 42, 165015, \dodoi{10.1088/1361-6382/adf58b}

\bibitem[{Zevin {et~al.}(2017)}]{Zevin:2016qwy}
Zevin, M., {et~al.} 2017, Class. Quant. Grav., 34, 064003,
  \dodoi{10.1088/1361-6382/aa5cea}

\bibitem[{Zevin {et~al.}(2024)}]{Zevin:2023rmt}
---. 2024, Eur. Phys. J. Plus, 139, 100,
  \dodoi{10.1140/epjp/s13360-023-04795-4}

\end{thebibliography}
\end{document}